\documentclass[apj]{emulateapj}
\newcommand{\DDir}{\relax{D\kern-.7em{/}}}

\newcommand{\xra}{\xrightarrow}

\newcommand{\be}{\begin{equation}}
\newcommand{\ee}{\end{equation}}
\newcommand{\bea}{\begin{equation*}}
\newcommand{\eea}{\end{equation*}}

\newcommand{\nin}{\relax{\in\kern-.8em{/}}}

\newcommand{\TeV}{\mbox{ TeV}}

\usepackage{amsmath}

\begin{document}

\title{Nonthermal emission from clusters of galaxies}

\author{Doron Kushnir\altaffilmark{1} and
Eli Waxman\altaffilmark{1}}
\altaffiltext{1}{Physics Faculty,
Weizmann Institute of Science, Rehovot, Israel}

\begin{abstract}

We show that the spectral and radial distribution of the nonthermal emission of massive, $M\gtrsim10^{14.5}M_\sun$, galaxy clusters may be approximately described by simple analytic expressions, which depend on the cluster thermal X-ray properties and on two model parameter, $\beta_{\textrm{core}}$ and $\eta_e$. $\beta_{\textrm{core}}$ is the ratio of the cosmic-ray (CR) energy density (within a logarithmic CR energy interval) and the thermal energy density at the cluster core, and $\eta_{e(p)}$ is the fraction of the thermal energy generated in strong collisionless shocks, which is deposited in CR electrons (protons). Using a simple analytic model for the evolution of intra-cluster medium CRs, which are produced by accretion shocks, we find that $\beta_{\textrm{core}}\simeq\eta_{p}/200$, nearly independent of cluster mass and with a  scatter $\Delta\ln\beta_{\textrm{core}}\simeq1$ between clusters of given mass. We show that the hard X-ray (HXR) and $\gamma$-ray luminosities produced by inverse Compton scattering of CMB photons by electrons accelerated in accretion shocks ({\it primary electrons}) exceed the luminosities produced by secondary particles (generated in hadronic interactions within the cluster) by factors $\simeq500(\eta_e/\eta_p)(T/10{\rm keV})^{-1/2}$ and $\simeq150(\eta_e/\eta_p)(T/10{\rm keV})^{-1/2}$ respectively, where $T$ is the cluster temperature. Secondary particle emission may dominate at the radio and very high energy ($\gtrsim1$~TeV) $\gamma$-ray bands. Our model predicts, in contrast with some earlier work, that the HXR and $\gamma$-ray emission from clusters of galaxies are extended, since the emission is dominated at these energies by primary (rather than by secondary) electrons. Our predictions are consistent with the observed nonthermal emission of the Coma cluster for $\eta_p\sim\eta_e\sim0.1$. The implications of our predictions to future HXR observations (e.g. by NuStar, Simbol-X) and to (space/ground based) $\gamma$-ray observations (e.g. by Fermi, HESS, MAGIC, VERITAS) are discussed. In particular, we identify the clusters which are the best candidates for detection in $\gamma$-rays. Finally, we show that our model's results agree with results of detailed numerical calculations, and that discrepancies between the results of various numerical simulations (and between such results and our model) are due to inaccuracies in the numerical calculations.

\end{abstract}

% -------------------------- End of abstract -----------------------

\keywords{ acceleration of particles - galaxies: clusters: general -
radiation mechanisms: nonthermal - X-rays: general}

% -----------------------------------------------------------------------
% --------------------------  Sec 1: INTRODUCTION -----------------------
% -----------------------------------------------------------------------

\section{Introduction}
\label{sec:Introduction}

A nonthermal emission component is observed in several clusters of
galaxies. In most cases nonthermal radio emission is observed \citep{feretti2005nte},
and in some cases a nonthermal hard X-ray (HXR) emission component is also observed
\citep[for review, see][]{rephaeli2008npc}. The radio emission is interpreted as synchrotron radiation, thereby suggesting that relativistic electrons and magnetic fields are present in the intracluster medium (ICM).

Several models for the synchrotron emission have been presented in the literature. These models differ in the assumptions regarding the origin of the emitting electrons. In some models the emitting electrons are secondary electrons and positrons that were generated by p-p interactions of a cosmic ray (CR) proton population with the ICM \citep[e.g.][]{dennison1980frh,blasi1999crr}, while in others the high energy of the emitting electrons is acheived by turbulent acceleration of a preexisting ICM population of nonthermal "seed electrons" \citep[secondary or otherwise, e.g.][]{brunetti2001prc,petrosian2001one,brunetti2005arr,cassano2005cmn,cassano2007nsr,brunetti2008gre}. Various sources of CR protons and electrons were considered in the literature, including   active galactic nuclei \citep[e.g.][]{katz1976oxr,fabian1976gcs,fujita2007nea}, dark matter bow shocks \citep[e.g.][]{bykov2000nec}, ram-pressure stripping of infalling galaxies \citep[e.g.][]{deplaa2006ces} and shock waves associated with the process of large scale structure (LSS) formation \citep[e.g.][]{loeb2000cgr,fujita2003nep,berrington2003npa,gabici2003nrc,brunetti2004arr,inoue2005hxr}.
In this paper we consider CRs produced by LSS shocks, and derive predictions for the nonthermal cluster emission they generate. Detailed analysis, based on the predictions of this paper, of the nonthermal radio and HXR emission from a sample of clusters suggests that CR acceleration in LSS shocks is the main source of cluster CRs \citep{kushnir2009mfc,kushnir2009hxr}.

The following point regarding the general applicability of our results should be emphasized here. The simple analytic expressions we derive (in \S~\ref{sec:simple}) give the non-thermal emission from secondary particles (produced by CR interactions at the cluster core) as a function of $\beta_{\rm core}$ and of the observed thermal X-ray properties of the cluster. These expressions are valid, for a given value of $\beta_{\rm core}$, regardless of the origin of the CRs residing at the cluster core. As shown in \citep{kushnir2009mfc}, cluster observations already allow one to estimate both the average value of $\beta_{\rm core}$ and its scatter, $\Delta\ln\beta_{\rm core}$. The values derived from observations are naturally explained within the frame of the model presented here, where cluster CRs are produced by LSS shocks, and difficult to explain in other models. Moreover, since alternative models for the origin of cluster CRs must reproduce the observationally inferred values of $\beta_{\rm core}$ and $\Delta\ln\beta_{\rm core}$, the non-thermal secondary emission predicted by such models would be similar to that derived here.

There are two primary populations of LSS shocks, which differ in their Mach numbers \citep[see e.g.][]{ryu2003csw,pfrommer2006dsw,skillman2008csa}: High Mach number shocks generated by the accretion of gas onto filaments and clusters ({\it accretion shocks}), and low Mach number shocks produced by halo mergers ({\it merger shocks}). The hypothesis that accretion shocks generate high energy CRs is supported by their resemblance to collisionless non-relativistic shocks in the interstellar medium \citep[see e.g.][]{keshet2003gri}, which are generally known to accelerate a power-law distribution of high energy particles \citep{blandford1987paa}.

Accretion shocks are expected to accelerate electrons to $\textrm{TeV}$ energies. Such electrons, to which we refer as {\it primary electrons}, lose their energy by inverse Compton (IC) scattering of CMB photons on a time scale which is much shorter then the cluster dynamical time, leading to HXR and $\gamma$-ray emission at the vicinity of the shocks
\citep[see e.g.][]{dar1995ohe,colafrancesco1998cgd,loeb2000cgr,waxman2000frb,totani2000fcg,kawasaki2002pcb}. Accretion shocks are also expected to accelerate protons to high energies. The protons, which do not lose their energy on a cluster dynamical time, may be coupled to the thermal plasma by magnetic fields and be confined within the cluster volume over cosmological times. Inelastic collisions of these CR protons with thermal ICM protons would produce high energy $\gamma$-rays through the decay of neutral pions, and high energy secondary electrons and positrons through the decay of charged pions \citep{dennison1980frh,volk1996nec,berezinsky1997cgs}. The secondary nonthermal emission is dominated by the cluster's core, where the ICM and CR density is highest.

To date, nonthermal emission from
accretion shocks has not yet been reliably observed. Furthermore, no cluster of
galaxies has so far been firmly detected in $\gamma$-rays
\citep{reimer2003eul} or in very high energy (VHE, $\gtrsim1$~TeV) $\gamma$-rays \citep{perkins2006tgr,domainko2007hog,perkins2008voc} \citep[see, however, the stacking analysis of][which revealed a low significance excess of $\gamma$-ray emission associated with Abell clusters]{scharf2002sdg}.
As we discuss in \S~\ref{sec:discussion}, future HXR observations (e.g. NuStar, Simbol-X) and (space/ground based) $\gamma$-ray observations (e.g. Fermi, HESS, MAGIC, VERITAS) are expected to provide unambiguous detection of nonthermal emission from accretion shocks.

In order to determine the nonthermal relativistic particle population in clusters, one has to follow the hydrodynamic history of the gas confined within these structures. Nonthermal relativistic particles (CRs) are produced in accretion and
merger shocks, and the CR energy density is modified by adiabatic expansion/compression of the gas, and by energy loss due to emission of radiation and due to inelastic nuclear collisions. The CR population in clusters may also be affected by diffusion. Diffusion of CRs on scales $\sim100$~kpc would reduce their density at the cores of clusters and hence the secondary emission. The effects of diffusion are difficult to estimate theoretically, mainly due to our ignorance regarding the structure of magnetic fields within the ICM. However, cluster radio observations imply that diffusion does not significantly affect the secondary population, and that the diffusion time of 100~GeV CRs over scales $\gtrsim100$~kpc within the ICM is not short compared to the Hubble time \citep{kushnir2009mfc}. We therefore neglect diffusion in the present analysis.

The evolution of CRs in large-scale structures was investigated numerically in several earlier studies \citep{keshet2003gri,miniati2003nmg,pfrommer2007scr,pfrommer2008scr,pfrommer2008scrb}. All studies identified cluster cores and cluster accretion shocks as the main sources of nonthermal radiation. There are discrepancies between the results of some of the studies, regarding the particles (primaries or secondaries) that dominate the emission at different energies and at different spatial locations. Due to the complicated nature of numerical simulations it is difficult to trace the origin of the discrepancies, which may be largely due to the different values adopted in the simulations for the acceleration efficiencies in shocks. We will resolve here some of   these discrepancies by comparing the numerical results with the results of our analytic analysis.

Several earlier studies attempted to provide an analytic derivation of the nonthermal luminosity of galaxy clusters \citep{sarazin1999esp,fujita2001nea,berrington2003npa,gabici2003nrc,araudo2008ntp,murase2008cra}. These studies did not include a model for the evolution of the space and energy distributions of primary and secondary CRs due to both accretion/merger shocks and adiabatic expansion/compression. Moreover, they did not consider IC emission by primary electrons, which, as we show below, dominates the HXR and $\gamma$-ray emission.

We derive in this paper a simple analytic model describing the spatial and spectral distribution of the nonthermal emission produced by cluster CRs. We focus on massive clusters, $M\gtrsim10^{14.5}M_\sun$, since their nonthermal emission is strongest and may be detectable by current and upcoming experiments. We assume that accretion and merger shocks produce power-law momentum distributions of CRs, $d\ln n/d\ln p=-\alpha$, with $\alpha$ determined by the shock Mach number, $\mathcal{M}$, $\alpha=4/(1-\mathcal{M}^{-2})-2$ \citep{blandford1987paa}, and that strong ($\mathcal{M}^{2}\gg1$) shocks deposit a fraction $\eta_{p(e)}$ of the thermal energy they generate in CR protons (electrons). Our description of the accretion shocks, which dominate the CR generation (see \S~\ref{sec:simple}), is highly simplified. We assume that the clusters are spherical, that gas is accreted onto clusters in spherically symmetric flows at a rate of $\sim M/t_H$, where $t_H$ is the (instantaneous) Hubble time, and that accretion shocks are strong ($\mathcal{M}^2\gg1$). We show (in \S~\ref{sec:compare 3d}) that the results of numerical simulations are consistent with an average accretion rate (at low $z$) of $\approx 0.5M/t_H$.

The nonthermal emission is dominated by primary electrons near the accretion shocks and by secondaries at the cluster core. It is determined therefore mainly by two parameters, $\eta_{e}$ and $\beta_{\textrm{core}}$, the ratio of the CR energy density (within a logarithmic CR energy interval) and the thermal energy density at the cluster core. We first derive in \S~\ref{sec:simple} simple analytic expressions for the nonthermal emission as function of $\eta_{e}$ and $\beta_{\textrm{core}}$, and of the thermal X-ray emission properties of the cluster (the charged secondary, $e^\pm$, emission, which may dominate at radio wavelengths, depends also on the strength of the ICM magnetic field). We also show, based on simple physical arguments and crude approximations, that $\beta_{\textrm{core}}\simeq\eta_{p}/100$ should be expected in massive clusters \citep[see also][]{pfrommer2007scr,jubelgas2008crf}. The results of our simple model are shown to be consistent with those of detailed numerical simulations in \S~\ref{sec:compare 3d}, with the exception of deviations which are due to inaccuracies of the numerical calculations. In \S~\ref{sec:compare observations} we compare our model predictions with the observed nonthermal emission of the Coma cluster. The extended HXR emission of Coma is shown to be consistent with our model predictions, and the HXR and radio luminosities are shown to be consistent with our model predictions for $\eta_{p}\sim\eta_{e}\sim0.1$.

Some limitations of our simplified model should be highlighted. Our model does not capture temporal fluctuations in the mass accretion rate and deviations of the accretion flow from spherical symmetry. Since the cooling time of \textit{primary electrons} is short compared to the cluster dynamical time, these effects will lead to temporal fluctuations and to deviations from spherical symmetry of the HXR and gamma-ray emission. Our model approximately describes therefore only the temporally and azimuthally averaged emission of primary electrons. Dedicated numerical simulations that constrain the temporal and spatial fluctuations of the accretion rate will be useful for analyzing upcoming observations (as discussed in detail in \S~\ref{sec:discussion}). The emission produced by \textit{secondaries}, on the other hand, is not sensitive to the spatial and temporal fluctuations in the accretion of gas. This is due to the fact that the cooling time of CR protons, which produce the secondaries, is long (of the order of the Hubble time). They therefore accumulate in the cluster over its age, and their density at the cluster core, which dominates the secondary emission, is not sensitive to fluctuations in the accretion rate.

In \S~\ref{sec:dynamics} we examine the validity of the relation $\beta_{\textrm{core}}\simeq\eta_{p}/100$. We determine the value of $\beta_{\textrm{core}}/\eta_{p}$, its scatter and its dependence on cluster mass using a simple toy model for the evolution of cluster CRs. In order to describe the effects of mergers, we construct (in \S~\ref{sec:acc_and_merger}) the merger history of clusters using the scheme of \citet{lacey1993mrh}, and construct (in \S~\ref{sec:merger}) a simple model describing the shocks and the adiabatic compression/expansion of the gas induced by mergers, which captures the main features of 3D numerical merger simulations \citep[e.g.][]{mccarthy2007msh}. Using the simple model of \S~\ref{sec:acc_and_merger} and \S~\ref{sec:merger} for   the evolution of ICM gas, we construct in \S~\ref{sec:cr dynamics} a model for the evolution of the CR population. We assume that the effects of diffusion are small, and that CRs are advected with the gas. The evolution of the CR population is followed taking into account energy losses due to Coulomb and inelastic nuclear collisions and assuming that the relativistic particles behave as an ideal gas with an adiabatic index of $4/3$ \citep[see e.g.][]{ensslin2007crp}. The results of this model slightly modify out crude estimation to $\beta_{\textrm{core}}\simeq\eta_{p}/200$, with small scatter and weak dependence on cluster mass.

One limitation of the toy model for ICM evolution is that it includes a description of entropy changes that are driven only by gravity (accretion and merger shocks). It does not include a description of a possible increase of the entropy at high redshift by non-gravitational processes \citep[such as supernovae, star formation and galactic winds, e.g.][]{kaiser1991ecg,evrard1991exr,navarro1995sxr,cavaliere1997tlt,balogh1999phi,ponman1999tti}, and of the decrease of entropy due to cooling at the centers of massive clusters \citep[see e.g.][]{fabian1994cfc,voigt2004tcr}. As explained in \S~\ref{sec:entropy_changes}, for massive clusters, $M\gtrsim10^{14.5}M_\sun$, these effects are unlikely to modify significantly the main conclusion of \S~\ref{sec:cr dynamics}, $\beta_{\textrm{core}}\simeq\eta_{p}/200$, and are therefore unlikely to significantly modify the predicted properties of the nonthermal emission.

In \S~\ref{sec:discussion} we explain how our model predictions may be tested, and how the values of the parameters $\eta_{p}$ and $\eta_{e}$ may be calibrated, using a controlled sample of clusters observed in radio and HXRs. We also discuss our results' implications for future $\gamma$-ray observations with space-borne and ground based telescopes. In particular, we identify the clusters which are the best candidates for detection in $\gamma$-rays, and the best candidates for detection of emission from pion decays.

Throughout, a $\Lambda$CDM cosmological model is assumed with a Hubble constant $H_{0}=70h_{70}\,\textrm{km}\,\textrm{s}^{-1}\,\textrm{Mpc}^{-1}$, $\Omega_{m}=0.23$, $\Omega_{b}=0.039$, $\Omega_{\Lambda}=1-\Omega_{m}$, and $\sigma_{8}=0.9$

% --- Sec 2: Simple model -------------
% -----------------------------------------------------------------------

\section{A simple analytic model for the nonthermal emission}
\label{sec:simple}

In this section we derive the spectral and radial distribution of the nonthermal emission produced by ICM CRs in massive clusters, $M\gtrsim10^{14.5}M_\sun$. We assume that the fraction $\beta_{\textrm{core}}$ of plasma energy carried by CR protons at the central regions of clusters, which dominate the emission from secondary particles, is nearly independent of cluster mass and has a small scatter. We show below that the secondary nonthermal emission of the clusters is determined by their thermal X-ray emission properties and by the value of $\beta_{\textrm{core}}$ (the secondary $e^\pm$ emission depends also on the strength of the magnetic field within the core). The primary nonthermal emission depends on the thermal X-ray properties and on $\eta_{e}$. This allows us to derive simple analytic expressions for the spectral and spatial distribution of the nonthermal emission. The analytic expressions show explicitly the dependence of the nonthermal emission on model parameters (e.g., the influence of a scatter in $\beta_{\textrm{core}}$ is easily inferred).

Before going into the details of the model, let us briefly explain, using simple arguments, why we expect $\beta_{\textrm{core}}\simeq\eta_p/100$ \citep[see also][]{pfrommer2007scr,jubelgas2008crf}. As we show in section \S~\ref{sec:dynamics}, the generation of CRs is dominated by accretion shocks. Since accretion shocks are characterized by high Mach numbers, we expect them to produce a flat CR energy spectrum,
\begin{eqnarray}\label{eq:flat spectrum}
dn_{\rm CR}/d\varepsilon\propto \varepsilon^{-2}.
\end{eqnarray}
For this distribution, the post-shock energy density of CR protons within a logarithmic proton energy interval (around $\varepsilon$) is related to the fraction $\eta_p$ of the post-shock thermal plasma energy deposited in CR protons by
\begin{eqnarray}\label{eq:beta def}
\beta_{\textrm{CR},p}(\varepsilon)\equiv\frac{ \varepsilon^{2} dn_{{\rm CR},p}/{d\varepsilon}}{\varepsilon_{\textrm{gas}}}\approx
\frac{\eta_{p}}{\ln[p_{\textrm{max}}/(1{\rm GeV}/c)]}.
\end{eqnarray}
Here $\varepsilon_{\textrm{gas}}$ is the post-shock thermal plasma energy density and $p_{\textrm{max}}$ is the maximal momentum of the CR proton spectrum respectively. The maximal momentum is determined by comparing the acceleration time $r_{\textrm{L,p}}c/v_{\textrm{sh}}^2\simeq2.9\cdot10^{6}\gamma_{7}(B_{-7}T_{1})^{-1}\,\textrm{yr}$ (where $v_{\textrm{sh}}=8\sqrt{3T/2m_{p}}/3$ is the shock velocity, $r_{\textrm{L,p}}=10^{2}\gamma_{7}/B_{-7}\,\textrm{pc}$ is the Larmor radius of the proton, $\gamma_{7}\equiv\gamma/10^{7}$ is the Lorentz factor of the proton, $B_{-7}\equiv B/0.1\,\mu\textrm{G}$ and $T_{1}\equiv T/10\,\textrm{keV}$ is the cluster temperature) to the typical cluster dynamical time, $t_{\rm dyn}\sim10^{9}\,\textrm{yr}$, which gives $p_{\textrm{max}}\sim10^{18}\,\textrm{eV}/c$. Since $\ln[p_{\textrm{max}}/(1{\rm GeV}/c)]\simeq 20$, $\beta_{\textrm{CR},p}\simeq\eta_{p}/20$ right behind the accretion shock.

As we show in section \S~\ref{sec:dynamics}, merger shocks do not significantly affect the CR population. This, and the long (inelastic nuclear collision) cooling time of CR protons,
\begin{eqnarray}\label{eq:p cool}
t_{\textrm{pp}}\simeq\left(\sigma_{\textrm{pp}}^{\textrm{inel}}cn\right)^{-1}
\simeq2.6\cdot10^{10}\left(\frac{n}{10^{-3}\,\textrm{cm}^{-3}}\right)^{-1}\,\textrm{yr},
\end{eqnarray}
where $\sigma_{\textrm{pp}}^{\textrm{inel}}\simeq40\,\textrm{mb}$ and $n$ is the gas number density,
implies that the energy density of CR protons produced by the accretion shock is later affected mainly by adiabatic expansion and compression. The difference between the adiabatic indices of the relativistic CRs ($4/3$) and non-relativistic thermal plasma ($5/3$) implies $\beta_{\textrm{CR},p}\propto\rho_{\textrm{gas}}^{-1/3}$, where $\rho_{\rm gas}$ is the gas density. The ratio between the mean gas density in the core and the gas density behind the accretion shock is typically $\sim10^{2}$, implying that the value of $\beta_{\textrm{CR},p}$ at the cluster core is typically expected to be $\beta_{\textrm{core}}\simeq\eta_{p}/100$.

We have ignored in the preceding discussion the fact that different mass elements were accreted at different redshifts, hence at a different densities, which implies that their density compression factor is not given directly by the ratio of densities at the cluster core and at the accretion shock. Moreover, major cluster mergers mix the gas and heat it with weak shocks, such that some deviations from the proposed simple relation are expected. We present in \S~\ref{sec:dynamics} a simple toy model for the evolution of cluster CRs, which examines the significance of these effects. The model yields $\beta_{\textrm{core}}\simeq\eta_{p}/200$, nearly independent of cluster mass and with small scatter among clusters of given mass. We also discuss in \S~\ref{sec:dynamics} the effects of entropy increase at high redshift (by non-gravitational processes) and of entropy decrease due to cooling at the centers of massive clusters. We argue that these effects are unlikely to modify significantly the predicted properties of the nonthermal emission from massive clusters.

We derive below analytic expressions for the luminosity and surface brightness produced by different emission mechanisms. For convenience, the details of the calculations are given in \S~\ref{sec:appendix A}, and only the main results are given in \S~\ref{sec:p rad}--\S~\ref{sec:rad sum}. We note that since the emission from hadronic processes is concentrated at the cluster core, while the emission from primary electrons originates from a thin layer around the accretion shock, the surface brightness profiles produced by primary and secondary particles are different. One may therefore distinguish between the two sources of nonthermal radiation by examining the radial dependence of the nonthermal emission surface brightness. We first describe in \S~\ref{sec:properties} the thermal emission properties of galaxy clusters, based on which the nonthermal emission properties are later derived. In \S~\ref{sec:p rad} and \S~\ref{sec:sec rad} we summarize our model results for the emission from neutral ($\pi^{0}$) and charged ($e^\pm$) products of inelastic nuclear collisions of CR protons. In \S~\ref{sec:e rad} we summarize our model results for the (IC and synchrotron) emission from primary CR electrons. Motivated by the results of \citet{kushnir2009mfc}, who constrain $\beta_{\textrm{core}}$ based on radio emission, and by the results of \citet{kushnir2009hxr}, who constrain $\eta_e$ based on HXR observations, numerical results are given using a normalization of $\beta_{\textrm{core}}=10^{-4}$ and $\eta_e=0.01$.

The analytic results of \S~\ref{sec:p rad}--\S~\ref{sec:e rad} are obtained by using simple approximations for the spectrum of secondaries produced and for the IC and synchrotron spectra (the details are given in \S~\ref{sec:appendix A}). In \S~\ref{sec:rad sum} we compare the results of \S~\ref{sec:p rad}--\S~\ref{sec:e rad} to those obtained using a more accurate parametrization of the spectrum of p-p secondaries \citep[following][]{kamae2006pan}, and using the exact formulae for IC and synchrotron emission \citep[following][]{blumenthal1970bsr}. We find that the deviations are small.

The charged secondary emission depends on the the strength of the magnetic field in the cluster core. It is given in \S~\ref{sec:sec rad} in terms of $B/B_{\textrm{CMB}}$, the ratio of the magnetic field to $B_{\textrm{CMB}}$, defined as the magnetic field for which the magnetic energy density equals the CMB energy density,
\begin{eqnarray}\label{eq:B eq ucmb}
B_{\textrm{CMB}}\equiv(8\pi aT_{CMB}^4)^{1/2}=3.2(1+z)^{2}\,\mu \textrm{G}.
\end{eqnarray}
Since the secondary $e^\pm$ emission is dominated at high energy, $\gtrsim1$~eV, by secondary $\pi^0$ decay and by IC emission of primary electrons, the uncertainty in the value of $B$ affects only the predicted radio emission, which is not the main focus of the current paper. The detailed discussion of radio emission from clusters given in \citet{kushnir2009mfc} indicates that $B$ is within the range of $\sim1\mu$G to $\sim10\mu$G.

\subsection{Clusters of galaxies: thermal emission properties}
\label{sec:properties}

The thermal X-ray emission of clusters is usually characterized by 4 parameters: $L_{X}$, the bolometric luminosity integrated out to some radius $r_{X}$ from the cluster center; $T$, the cluster temperature (weighted averaged over the cluster emission); $\beta$ and $r_{c}$, which define the radial surface brightness profile, and hence the radial density profile \citep{cavaliere1976xrh,gorenstein1978sxr,jones1984scg},
\begin{equation}\label{eq:beta model}
\rho_{\textrm{gas}}(r)=\rho_{0}\left(1+\frac{r^{2}}{r_{c}^{2}}\right)^{-(3/2)\beta}.
\end{equation}
Using the HIFLUCGCS sample of the X-ray-brightest galaxy clusters of \citet{reiprich2002mfx}, which is based on the ROSAT All-Sky X-Ray Survey, we derive the following correlations:
\begin{eqnarray}\label{eq:power law form}
L_{X}&=&L_{X0}h_{70}^{-2}T_{1}^{\alpha_{L}}, \nonumber \\ r_{c}&=&r_{c0}h_{70}^{-1}T_{1}^{\alpha_{r}},
\end{eqnarray}
where $L_{X0}=2.76\cdot10^{45}\,\textrm{erg}\,\textrm{s}^{-1}$, $\alpha_{L}=2.56$, $r_{c0}=223\,\textrm{kpc}$ and $\alpha_{r}=1.32$. \citep[Note, that various authors obtain values of $\alpha_{L}$ in the
range $\approx 2.5-3$,
e.g.][]{markevitch1998lxt,arnaud1999tlx,reiprich2002mfx}.

In what follows, we give expressions for the nonthermal emission as function of $T,\ \beta,\ r_c,$ and $L_X$. We also give expressions which depend on $T$ and $\beta$ only, using the correlations given in eq.~(\ref{eq:power law form}). The following cautionary note should be made in this context. Our demonstration in \S~\ref{sec:dynamics}, that $\beta_{\textrm{core}}\simeq\eta_{p}/200$ with weak dependence on cluster mass, is based on a model that does not include entropy changes not induced by LSS shocks. As explained there, and also in \S~\ref{sec:Introduction}, early (high redshift) entropy injection (which does not significantly affect massive clusters) may modify the value of $\beta_{\textrm{core}}$ for low mass clusters, thus introducing at low $T$ an   additional, implicit, dependence on $T$ through $\beta_{\textrm{core}}$. Note, that such early entropy injections are believed to be responsible for the deviations at low $T$ of the observed correlations, eq.~(\ref{eq:power law form}), from the self-similar relations expected when entropy changes not induced by LSS shocks are neglected \citep[e.g. $L_{X}\propto T^{2}$, see][]{arnaud1999tlx}.

We define the cluster mass, $M_{200}$, as the mass contained within a radius $r_{200}$, within which the mean density is $200$ times the critical density, $\rho_{\textrm{crit}}$. This mass may be related to $T$ and $\beta$ by assuming the ICM plasma to be isothermal and in hydrostatic equilibrium. Under these assumptions, the gravitational mass is given by
\begin{equation}\label{eq:ravitational mass}
M(<r)=\frac{3\beta T}{\mu m_{p}G}\frac{r^{3}}{r_{c}^{2}+r^{2}}
\end{equation}
(the mean molecular weight is $\mu\simeq0.59$ for fully ionized plasma with hydrogen mass fraction $\chi=0.75$), and we obtain (assuming $r_{200}\gg r_{c}$)
\begin{eqnarray}\label{eq:r200,M200}
r_{200}&\simeq&\left(\frac{800\pi}{3}\rho_{\textrm{crit}}\right)^{-1/2} \left(\frac{3\beta T}{\mu m_{p}G}\right)^{1/2} \nonumber \\ &=&3.1\beta^{1/2}T_{1}^{1/2}h_{70}^{-1}\,\textrm{Mpc},\nonumber \\
M_{200}&\simeq&\left(\frac{800\pi}{3}\rho_{\textrm{crit}}\right)^{-1/2} \left(\frac{3\beta T}{\mu m_{p}G}\right)^{3/2} \nonumber \\ &=&3.5\cdot10^{15}\beta^{3/2}T_{1}^{3/2}h_{70}^{-1}M_{\odot}.
\end{eqnarray}
In addition to the assumption that the ICM is isothermal and in hydrostatic equilibrium, we have extrapolated here the surface brightness given by eq.~(\ref{eq:beta model}) out to $r_{200}$, which may be larger than $r_{X}$. This implies  $\rho\propto r^{-2}$ at large radii, in contrast with the $\rho\propto r^{-3}$ dependence expected at large $r$ \citep[e.g.][]{navarro1997udp}. The detailed discussion given in \citet{reiprich2002mfx} of the accuracy of cluster mass determination under these approximations shows that eq.~\eqref{eq:r200,M200} may overestimate $M_{200}$ by no more than $20\%$.

\subsection{$\pi^0$ decay emission}
\label{sec:p rad}

In this section we present our model results for the emission from $\pi^{0}$ decays. Since the CR proton spectrum is flat, the p-p $\gamma$-ray luminosity per logarithmic photon energy bin, $\nu L_{\nu}^{\textrm{pp}}$, is energy independent above $\sim0.1\,\textrm{GeV}$ (the threshold energy for pion production is $\varepsilon_{\textrm{th}}\simeq1.22\,\textrm{GeV}$ and the photon energy is $\sim0.1$ of the parent CR proton energy). The observed spectrum is expected to be cut off at high energy due to pair production interactions of the high energy photons with infra-red background photons. For nearby clusters, the $\gamma\gamma\rightarrow e^{+}e^{-}$ cutoff is expected at $\sim10\,\textrm{TeV}$ \citep{franceschini2008eoi}. The $\pi^{0}$ luminosity is given by (see eq.~\eqref{eq:Lpp app})
\begin{eqnarray}\label{eq:Lpp}
\nu L_{\nu}^{\textrm{pp}}&\simeq&  1.3\cdot10^{41}\beta_{\textrm{core},-4} T_{1}^{1/2} \nonumber \\
&\times&\left(\frac{L_{X}}{h_{70}^{-2} 3\cdot10^{45}\,\textrm{erg}\,\textrm{s}^{-1}}\right)\,\textrm{erg}\,\textrm{s}^{-1},
\end{eqnarray}
where $\beta_{\textrm{core},-4}=\beta_{\textrm{core}}/10^{-4}$.
$\nu L_{\nu}^{\textrm{pp}}$ depends linearly on $L_{X}$, since both hadronic emission and thermal bremsstrahlung emission depend on density squared \citep{katz08}. Using the correlations eq.~\eqref{eq:power law form} we have
\begin{eqnarray}\label{eq:Lpp corr}
&\nu L_{\nu}^{\textrm{pp}}\simeq 1.2\cdot10^{41}\beta_{\textrm{core},-4} T_{1}^{3.06} \,\textrm{erg}\,\textrm{s}^{-1}.
\end{eqnarray}

The p-p $\gamma$-ray surface brightness above some energy $\varepsilon_{\nu,\textrm{min}}$ at some distance $\bar{r}\equiv r/r_{c}$ from the cluster center is given by \citep[for $\beta>0.5$, see eq.~\eqref{eq:Spp app} and][]{Sarazin1977xle}
\begin{eqnarray}\label{eq:Spp}
S_{\nu>\nu_{\textrm{min}}}^{\textrm{pp}} (\bar{r})&\simeq&
5.5\cdot10^{-7}\left(3\beta-\frac{3}{2}\right) \beta_{\textrm{core},-4} T_{1}^{1/2} \nonumber \\
&\times&\left(\frac{L_{X}}{h_{70}^{-2} 3\cdot10^{45}\,\textrm{erg}\,\textrm{s}^{-1}}\right) \left(\frac{r_{c}}{h_{70}^{-1}\,200\,\textrm{kpc}}\right)^{-2} \nonumber \\
&\times& \left(\frac{\max(\varepsilon_{\nu,\textrm{min}},0.1\varepsilon_{\textrm{th}})}{10\,\textrm{GeV}}\right)^{-1}
 \nonumber\\ &\times&
\left(1+\bar{r}^{2}\right)^{-3\beta+1/2} \,\textrm{ph}\,\textrm{cm}^{-2}\,\textrm{s}^{-1}\,\textrm{sr}^{-1}.
\end{eqnarray}
Using the correlations eq.~\eqref{eq:power law form} we have
\begin{eqnarray}\label{eq:Spp corr}
S_{\nu>\nu_{\textrm{min}}}^{\textrm{pp}} (\bar{r})&\simeq&  4.1\cdot10^{-7}\left(3\beta-\frac{3}{2}\right)\beta_{\textrm{core},-4}T_{1}^{0.42} \nonumber\\
&\times& \left(\frac{\max(\varepsilon_{\nu,\min},0.1\varepsilon_{\textrm{th}})}{10\,\textrm{GeV}}\right)^{-1} \nonumber\\  &\times& \left(1+\bar{r}^{2}\right)^{-3\beta+1/2} \,\textrm{ph}\,\textrm{cm}^{-2}\,\textrm{s}^{-1}\textrm{sr}^{-1}.
\end{eqnarray}

Note, that in the derivation of the surface brightness given above we have neglected the variation of $\beta_{\rm CR,p}$ with radius, and used $\beta_{\rm CR,p}=\beta_{\textrm{core}}$. This is justified since the surface brightness $S$ is $\propto\beta_{\rm CR,p}\rho_{\rm gas}^{(6\beta-1)/3\beta}$, and, as we show in \S~\ref{sec:dynamics}, $\beta_{\rm CR,p}\propto\rho_{\rm gas}^{-1/3}$. This implies that the variation of the surface brightness is dominated by the $\rho_{\rm gas}^{(6\beta-1)/3\beta}$ term (for typical $\beta$ values, which are in the range of 0.5 to 1).

\subsection{Emission from charged secondaries}
\label{sec:sec rad}

p-p collisions also produce secondary electrons and positrons, which cool by emitting synchrotron radiation and by IC scattering of CMB photons. Since variations in the secondary injection and in the cluster magnetic field occur on time scales similar to, or larger than, the dynamical time scale of the cluster, $t_{\rm dyn}\sim1\,\textrm{Gyr}$, the distribution of secondaries is in a steady state at high energies, at which the secondaries lose all their energy to radiation on a time scale short compared to the cluster dynamical time. For magnetic field values of $1-10\,\mu \textrm{G}$, the Lorentz factor of secondaries with cooling time, $6\pi m_{e}c/(B^{2}+B_{\rm CMB}^2)\sigma_{T}\gamma$, which equals the dynamical time, is $200\lesssim\gamma_{\textrm{cool}}\lesssim2000$.

The luminosity due to IC scattering of CMB photons by secondaries is given by (see eq.~\eqref{eq:sec IC app})
\begin{eqnarray}\label{eq:sec IC}
\nu L_{\nu}^{\textrm{IC},e^{\pm}}&\simeq&
3.3\cdot10^{40}\frac{B_{\textrm{CMB}}^{2}}{B_{\textrm{CMB}}^{2}+B^{2}}\beta_{\textrm{core},-4}T_{1}^{1/2} \nonumber \\
&\times&\left(\frac{L_{X}}{h_{70}^{-2} 3\cdot10^{45}\,\textrm{erg}\,\textrm{s}^{-1}}\right) \,\textrm{erg}\,\textrm{s}^{-1}.
\end{eqnarray}
Eq.~(\ref{eq:sec IC}) holds for photon energies in the range $\gamma_{\textrm{cool}}^{2}3T_{\textrm{CMB}}(1+z)^4<\varepsilon_{\textrm{ph}} <\gamma_{\textrm{max}}^{2}3T_{\textrm{CMB}}(1+z)^4$, where $\gamma_{\textrm{max}}\simeq0.1\varepsilon_{\textrm{max}}/m_{e}c^{2}$ ($\varepsilon_{\textrm{max}}$ is the maximal energy of the CR protons, and the secondary energy is $\sim0.1$ of the parent proton energy). The surface brightness is given by (see eq.~\eqref{eq:SICepm app})
\begin{eqnarray}\label{eq:SICepm}
S_{\nu>\nu_{\min}}^{\textrm{IC},e^{\pm}} (\bar{r})&\simeq&
1.4\cdot10^{-7} \left(3\beta-\frac{3}{2}\right) \frac{B_{\textrm{CMB}}^{2}}{B_{\textrm{CMB}}^{2}+B^{2}} \nonumber \\
&\times& \beta_{\textrm{core},-4} T_{1}^{1/2} \left(\frac{L_{X}}{h_{70}^{-2} 3\cdot10^{45}\,\textrm{erg}\,\textrm{s}^{-1}}\right) \nonumber\\
&\times& \left(\frac{r_{c}}{h_{70}^{-1}\,200\,\textrm{kpc}}\right)^{-2} \left(\frac{\varepsilon_{\nu,\min}}{10\,\textrm{GeV}}\right)^{-1} \nonumber\\
&\times&   \left(1+\bar{r}^{2}\right)^{-3\beta+1/2} \,\textrm{ph}\,\textrm{cm}^{-2}\,\textrm{s}^{-1}\,\textrm{sr}^{-1}.
\end{eqnarray}
Using the correlations eq.~\eqref{eq:power law form} we have
\begin{eqnarray}\label{eq:sec IC corr}
\nu L_{\nu}^{\textrm{IC},e^{\pm}}&\simeq& 3.0\cdot10^{40}\frac{B_{\textrm{CMB}}^{2}}{B_{\textrm{CMB}}^{2}+B^{2}} \nonumber\\ &\times& \beta_{\textrm{core},-4}T_{1}^{3.06} \,\textrm{erg}\,\textrm{s}^{-1},
\end{eqnarray}
and
\begin{eqnarray}\label{eq:SICepm corr}
S_{\nu>\nu_{\min}}^{\textrm{IC},e^{\pm}} (\bar{r})&\simeq&
1.0\cdot10^{-7}\left(3\beta-\frac{3}{2}\right) \frac{B_{\textrm{CMB}}^{2}}{B_{\textrm{CMB}}^{2}+B^{2}}  \nonumber\\
&\times&  \beta_{\textrm{core},-4}  T_{1}^{0.42} \left(\frac{\varepsilon_{\nu,\min}}{10\,\textrm{GeV}}\right)^{-1}  \nonumber\\
&\times& \left(1+\bar{r}^{2}\right)^{-3\beta+1/2} \,\textrm{ph}\,\textrm{cm}^{-2}\,\textrm{s}^{-1}\,\textrm{sr}^{-1}.
\end{eqnarray}
Note that although the $\gamma$-ray luminosity produced by $\pi^{0}$ decays is bigger above $\sim0.1\,\textrm{GeV}$ by a factor ${\simeq4(1+B^{2}/B_{\textrm{CMB}}^{2})}$ than that produced by charged secondaries (compare eq.~\eqref{eq:Lpp} and eq.~\eqref{eq:sec IC}), the secondaries' emission is the main hadronic emission mechanism below this energy.

For photons in the energy range $\gamma_{\textrm{cool}}^{2}\varepsilon_{0}<\varepsilon_{\textrm{ph}} <\gamma_{\textrm{max}}^{2}\varepsilon_{0}$, where $\varepsilon_{0}=3\hbar eB/2 m_{e}c$, the secondary synchrotron luminosity from is given by (see eq.~\eqref{eq:sec sync app})
\begin{eqnarray}\label{eq:sec sync}
\nu L_{\nu}^{\textrm{sync},e^{\pm}} &\simeq &  3.3\cdot10^{40}\frac{B^{2}}{B_{\textrm{CMB}}^{2}+B^{2}} \beta_{\textrm{core},-4}  T_{1}^{1/2}\nonumber \\
&\times& \left(\frac{L_{X}}{h_{70}^{-2} 3\cdot10^{45}\,\textrm{erg}\,\textrm{s}^{-1}}\right) \, \textrm{erg}\,\textrm{s}^{-1},
\end{eqnarray}
and the surface brightness is (see eq.~\eqref{eq:Ssyncepm app})
\begin{eqnarray}\label{eq:Ssyncepm}
S_{\nu}^{\textrm{sync},e^{\pm}} (\bar{r}) &\simeq&
13\cdot \left(3\beta-\frac{3}{2}\right) \frac{B^{2}}{B_{\textrm{CMB}}^{2}+B^{2}} \nonumber \\
&\times& \beta_{\textrm{core},-4} T_{1}^{1/2}
\left(\frac{L_{X}}{h_{70}^{-2} 3\cdot10^{45}\,\textrm{erg}\,\textrm{s}^{-1}}\right) \nonumber \\
&\times&\left(\frac{r_{c}}{h_{70}^{-1}\,200\,\textrm{kpc}}\right)^{-2} \left(\frac{\nu}{1.4\,\textrm{GHz}}\right)^{-1} \nonumber\\
&\times&  \left(1+\bar{r}^{2}\right)^{-3\beta+1/2}
\,\textrm{mJy}\,\textrm{arcmin}^{-2}.
\end{eqnarray}
Using the correlations eq.~\eqref{eq:power law form} we have
\begin{eqnarray}\label{eq:sec sync corr}
\nu L_{\nu}^{\textrm{sync},e^{\pm}}&\simeq & 3.0\cdot10^{40} \frac{B^{2}}{B_{\textrm{CMB}}^{2}+B^{2}} \beta_{\textrm{core},-4} \nonumber \\
&\times& T_{1}^{3.06}
\,\textrm{erg}\,\textrm{s}^{-1},
\end{eqnarray}
and
\begin{eqnarray}\label{eq:Ssyncepm corr}
S_{\nu}^{\textrm{sync},e^{\pm}} (\bar{r})&\simeq&
9.8\cdot \left(3\beta-\frac{3}{2}\right)  \frac{B^{2}}{B_{\textrm{CMB}}^{2}+B^{2}} \nonumber \\
&\times& \beta_{\textrm{core},-4} T_{1}^{0.42} \left(\frac{\nu}{1.4\,\textrm{GHz}}\right)^{-1}
 \nonumber\\
&\times& \left(1+\bar{r}^{2}\right)^{-3\beta+1/2}\,\textrm{mJy}\,\textrm{arcmin}^{-2}.
\end{eqnarray}

Note, that due to the uncertainties in $B$ and $t_{\rm dyn}$, which lead to uncertainty in $\gamma_{\textrm{cool}}$, our estimates of the emission from secondary electrons and positrons with $\gamma\sim\gamma_{\textrm{cool}}$ are uncertain.

\subsection{Primary electron emission}
\label{sec:e rad}

We now turn to estimating the radiation produced by primary electrons at the accretion shock. We assume that the clusters are spherical and that gas is accreted onto clusters in spherically symmetric flows at a rate $\dot{M}=f_{\textrm{inst}} M_{200}/t_H$. $f_{\textrm{inst}}$ is a dimensionless parameter of order unity, reflecting the temporal fluctuations of $\dot{M}/(M_{200}/t_H)$. As discussed in \S~\ref{sec:compare 3d}, 3D numerical simulations indicate that the average value of $f_{\textrm{inst}}$ is $\approx0.5$.

In order to calculate the surface brightness of the primary electrons' radiation, it is necessary to estimate the position of the accretion shock. We assume here that the accretion shock is located at $r\sim r_{200}$, since spherical collapse models predict a cluster virial density $<\rho_{\textrm{vir}}>\simeq178\rho_{\textrm{crit}}$ (for $\Omega_{m}=1$ and $\Omega_{\Lambda}=0$, with weak dependance on the background cosmology for $0.3\lesssim\Omega_{m}<1$). The validity of the simplifying assumptions described above is tested in \S~\ref{sec:compare 3d} by comparing our results to those of numerical 3D simulations. We find that the results obtained in this section are in good agreement with the results of detailed numerical simulations.

As in the case of the charged secondaries, we assume that the distribution of the primaries is in a steady state, since electrons (at the relevant energies) lose all their energy to radiation on a time scale short compared to the cluster dynamical time, $t_{\rm dyn}\sim1\,\textrm{Gyr}$. Unlike the secondary electrons and positrons, which lose energy through both IC and synchrotron emission, primary electrons lose their energy mainly by IC scattering of CMB photons, since the magnetic field at the accretion shock is expected to be weak, $\sim0.1\,\mu {\rm G}\ll B_{\rm CMB}$ \citep{waxman2000frb}. Thus the steady state assumption holds for primaries with Lorenz factors $\gamma>\gamma_{\textrm{cool}}\sim2000$. The maximal energy of the primary electrons is determined by equating the acceleration time, $r_{\textrm{L,e}}c/v_{\textrm{sh}}^2\simeq1.6\cdot10^{3}\gamma_{7}(B_{-7}T_{1})^{-1}\,\textrm{yr}$ (where $r_{\textrm{L,e}}=5.5\cdot10^{-2}\gamma_{7}/B_{-7}\,\textrm{pc}$ is the Larmor radius of the electron), to the IC cooling time, $6\pi m_{e}c/B^{2}_{\rm CMB}\sigma_{T}\gamma\simeq2.3\cdot10^{5}\gamma_{7}^{-1}(1+z)^{-4}\,\textrm{yr}$, which yields $\gamma_{\textrm{max}}\simeq1.2\cdot10^{8}\sqrt{B_{-7}T_{1}}(1+z)^{-2}$.

For photon energies $\gamma_{\textrm{cool}}^{2}3T_{\textrm{CMB}}(1+z)^4<\varepsilon_{\textrm{ph}} <\gamma_{\textrm{max}}^{2}3T_{\textrm{CMB}}(1+z)^4$, the primary IC luminosity is given, to very good accuracy, by (see eq.~\eqref{eq:shock IC app} and eq.~(\ref{eq:SICshock app}))
\begin{eqnarray}\label{eq:shock IC}
\nu L_{\nu}^{\textrm{IC,shock}}&\simeq&
1.8\cdot10^{43}\left(f_{\textrm{inst}}\eta_{e}\right)_{-2}\beta^{3/2} \nonumber \\
&\times& \left(\frac{f_{b}}{0.17}\right) T_{1}^{5/2} \bar{Z}(z) \,\textrm{erg}\,\textrm{s}^{-1}.
\end{eqnarray}
Here, $\left(f_{\textrm{inst}}\eta_{e}\right)_{-2}=f_{\textrm{inst}}\eta_{e}/10^{-2}$, $f_{b}=\Omega_{b}/\Omega_{m}$ and $\bar{Z}(z)\equiv(t_{H}H(z))^{-1}$ ($\bar{Z}(z)$ depends weakly on the assumed cosmology).
Assuming the emission originates from a thin layer with thickness $w$ behind the accretion shock, the surface brightness is given by
\begin{eqnarray}\label{eq:SICshock}
S_{\nu>\nu_{\min}}^{\textrm{IC,shock}}(r)&\simeq&
1.5\cdot10^{-7}\left(f_{\textrm{inst}}\eta_{e}\right)_{-2}\beta^{1/2} \nonumber \\ &\times&\left(\frac{f_{b}}{0.17}\right) T_{1}^{3/2} \left(\frac{\varepsilon_{\nu,\min}}{10\,\textrm{GeV}}\right)^{-1}\nonumber\\
&\times& \xi\left(r/r_{200},w/r_{200}\right) \nonumber\\
&\times&  \bar{Z}(z) h_{70}^2(z) \, \textrm{ph}\,\textrm{cm}^{-2}\,\textrm{s}^{-1}\,\textrm{sr}^{-1},
\end{eqnarray}
where
\begin{eqnarray}\label{eq:xi def}
\xi(x,y)=\left\{%
\begin{array}{ll}
    \frac{3\left(\sqrt{1-x^{2}}-\sqrt{(1-y)^{2}-x^{2}}\right)}{1-\left(1-y\right)^{3}}, & \hbox{$x\leq1-y$} \\
    \frac{3\sqrt{1-x^{2}}}{1-\left(1-y\right)^{3}}, & \hbox{$1-y<x<1.$} \\
\end{array}%
\right.
\end{eqnarray}
For $w\ll r_{200}$, $\xi(x,y)$ can be approximated in the regime $x<1-y$ by
\begin{eqnarray}\label{eq:xi apr}
\xi(x,y)\simeq\frac{1}{\sqrt{1-x^{2}}}.
\end{eqnarray}
The thickness $w$ of the emitting region is approximately given by the product of the cooling time of the emitting electrons, $t_{\textrm{cool}}$, and the velocity of the downstream fluid relative to the shock velocity, $u_{d}$. Since $t_{\textrm{dyn}}\sim r_{200}/u_{d}$, the approximation $w\ll r_{200}$ holds for $\gamma>\gamma_{\textrm{cool}}\sim2000$. Note that the rapid increase of the surface brightness near the accretion shock, inferred from eq.~\eqref{eq:xi apr}, is likely to be suppressed by small deviations from spherical symmetry.

Below $\sim1\,\textrm{TeV}$, the primary IC $\gamma$-ray luminosity is larger than the secondary $\gamma$-ray luminosity by a factor $\simeq150(f_{\textrm{inst}}\eta_{e})_{-2}(\beta_{\textrm{core}})_{-4}^{-1}\beta^{3/2}T_{1}^{-1/2}$  (compare eq.~\eqref{eq:shock IC} to the neutral component contribution eq.~\eqref{eq:Lpp corr}, which dominates the secondary emission at this energy range). The primary IC HXR luminosity is larger than the secondary HXR luminosity by a factor $\simeq600(f_{\textrm{inst}}\eta_{e})_{-2}(\beta_{\textrm{core}})_{-4}^{-1}\beta^{3/2}T_{1}^{-1/2}(1+B^{2}/B_{\rm CMB}^2)$ (compare eq.~\eqref{eq:shock IC} to the charged component contribution eq.~\eqref{eq:sec IC corr}, which dominates the secondary emission at this energy range). However, since the primary electrons' IC surface brightness is lowest at the cluster core, while the secondaries' emission surface brightness follows the thermal surface brightness (see e.g. eq.~\eqref{eq:Spp}), the surface brightness due to these two sources of radiation may be comparable at the cluster core.

For photon energies $\gamma_{\textrm{cool}}^{2}\varepsilon_{0}<\varepsilon_{\textrm{ph}} <\gamma_{\textrm{max}}^{2}\varepsilon_{0}$, the primary electrons' synchrotron luminosity is given by (see eq.~\eqref{eq:shock sync app})
\begin{eqnarray}\label{eq:shock sync}
\nu L_{\nu}^{\textrm{sync,shock}}&\simeq&
1.7\cdot10^{40}\left(f_{\textrm{inst}}\eta_{e}\right)_{-2} \beta^{3/2}\nonumber\\
&\times& \left(\frac{f_{b}}{0.17}\right) T_{1}^{5/2}
 B_{-7}^{2}  \nonumber\\
&\times& \bar{Z}(z)(1+z)^{-4} \,\textrm{erg}\,\textrm{s}^{-1}.
\end{eqnarray}
This luminosity is comparable to the luminosity produced by secondaries. However, the primary electron synchrotron surface brightness,
\begin{eqnarray}\label{eq:Ssyncshock}
S_{\nu}^{\textrm{sync,shock}}(r)&\simeq&
1.4\cdot10^{-2}\left(f_{\textrm{inst}}\eta_{e}\right)_{-2} \beta^{1/2}\nonumber \\
 &\times& \left(\frac{f_{b}}{0.17}\right) T_{1}^{3/2} \left(\frac{\nu}{1.4\,\textrm{GHz}}\right)^{-1}    \nonumber\\
&\times& B_{-7}^{2} \xi\left(r/r_{200},w/r_{200}\right)\bar{Z}(z) \nonumber\\
&\times&   h_{70}^2(z) (1+z)^{-4}\,\textrm{mJy}\,\textrm{arcmin}^{-2}
\end{eqnarray}
(see eq.~\eqref{eq:Ssyncshock app}), is negligible compared to that produced by secondaries (see eq.~\eqref{eq:Ssyncepm}).

\subsection{Summary of results}
\label{sec:rad sum}

\begin{figure}
\epsscale{1} \plotone{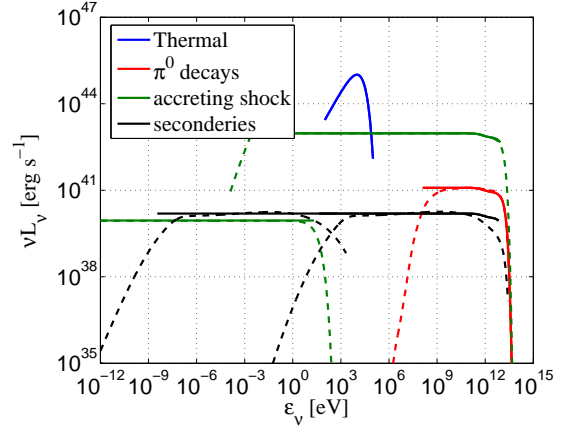} \caption{$\nu L_{\nu}$ as function of photon energy for a $T=10\,\textrm{keV}$, $\beta=2/3$ cluster, with $\beta_{\textrm{core}}=10^{-4}$, $f_{\textrm{inst}}\eta_{e}=0.01$, $B=B_{\textrm{CMB}}$ at the cluster core and $B=0.1\,\mu \textrm{G}$ at the accretion shock. Solid lines show the simple estimates given by eqs.~\eqref{eq:Lpp}, \eqref{eq:sec IC}, \eqref{eq:sec sync}, \eqref{eq:shock IC} and \eqref{eq:shock sync}, and dashed lines are obtained using the parametrization of \citet{kamae2006pan} for the secondary spectrum of p-p interactions, and the exact formulae for IC scattering and synchrotron emission given by \citet{blumenthal1970bsr}. Suppression of the flux by pair production (in interactions with IR background photons) was included, applying a suppression factor $\exp(-\tau_{\gamma\gamma})$ with pair production optical depth \citep[taken from][]{franceschini2008eoi} corresponding to a cluster at the distance of Coma. Note, that there is an artificial high energy cut-off in the secondary emission spectra calculated using the parametrization of \citet{kamae2006pan}, resulting from the upper limit on the proton energy used in this parametrization, $512\TeV$.
\label{fig:nuLnu_total}}
\end{figure}

Figure~\ref{fig:nuLnu_total} shows the nonthermal luminosity, $\nu L_{\nu}$, as function of photon energy for a $T=10\,\textrm{keV}$, $\beta=2/3$ cluster (with $L_{X}$ and $r_{c}$ determined by eq.~\eqref{eq:power law form}), with $\beta_{\textrm{core}}=10^{-4}$, $f_{\textrm{inst}}\eta_{e}=0.01$, $B=B_{\textrm{CMB}}$ at the cluster core and $B=0.1\,\mu \textrm{G}$ at the accretion shock (for our assumed background cosmology we have $\bar{Z}(z=0)\simeq0.97$). Solid lines show the simple estimates given above by eqs.~\eqref{eq:Lpp}, \eqref{eq:sec IC}, \eqref{eq:sec sync}, \eqref{eq:shock IC} and \eqref{eq:shock sync}, and the dashed lines are obtained using the parametrization of \citet{kamae2006pan} for the secondary spectrum of p-p interactions, and the exact formulae for IC scattering and synchrotron emission given by \citet{blumenthal1970bsr}. Suppression of the flux by pair production (in interactions with IR background photons) was included, applying a suppression factor $\exp(-\tau_{\gamma\gamma})$ with pair production optical depth \citep[taken from][]{franceschini2008eoi} corresponding to a cluster at the distance of Coma. Note, that there is an artificial high energy cut-off in the secondary emission spectra calculated using the parametrization of \citet{kamae2006pan}, resulting from the upper limit on the proton energy used in this parametrization, $512\TeV$. However, the secondaries' emission is always small in this range compared to other emission processes.

\begin{figure}
\epsscale{1} \plotone{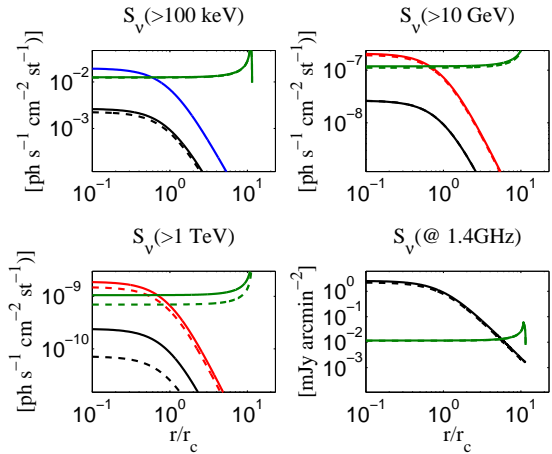} \caption{HXR, $\gamma$-ray and radio surface brightness as function of radius, given by eqs.~\eqref{eq:Spp}, \eqref{eq:SICepm}, \eqref{eq:Ssyncepm}, \eqref{eq:SICshock} and \eqref{eq:Ssyncshock}. Cluster parameters and line types are the same as in fig.~\ref{fig:nuLnu_total}. The thickness of the layer behind the accretion shock, from which the primary electron emission originates, is estimated to be $w\simeq u_{d}t_{\textrm{cool}}\ll r_{200}$. Note that the rapid increase of the surface brightness near the accretion shock is likely to be suppressed by small deviations from spherical symmetry.
\label{nu_gamma}}
\end{figure}

In fig.~\ref{nu_gamma} we show HXR, $\gamma$-ray and radio surface brightness profiles for the same cluster at various energies (see eqs.~\eqref{eq:Spp}, \eqref{eq:SICepm}, \eqref{eq:Ssyncepm}, \eqref{eq:SICshock} and \eqref{eq:Ssyncshock}). The thickness of the layer behind the accretion shock, from which the primary electron emission originates, is estimated to be $w\simeq u_{d}t_{\textrm{cool}}$. For the chosen cluster parameters, the dominant nonthermal emission process in the HXR ($>100\,\textrm{keV}$) band is IC emission from primary electrons at the accretion shock. At higher photon energies ($>10\,\textrm{GeV}$) the emission from pion decays at the core of the cluster becomes comparable to the IC emission from the accretion shock electrons. Note, that the IC emission from accretion shock electrons is uncertain at energies exceeding $\sim1$~TeV, due to the uncertainty in the maximal energy, $\sim$~few~TeV, to which electrons may be accelerated in such shocks. Thus, although at the figures shown the IC emission dominates the luminosity also at the highest energies, $>1$~TeV, pion decays may become the dominant emission process at these energies. At all energies, the contribution from secondary IC is small. However, the synchrotron emission of secondaries is the dominant emission process at the radio band.

% --------------------- End of section 2 --------------------------------

% -----------------------------------------------------------------------
% ---- Sec 3: Comparison to detailed numerical simulations --------------
% -----------------------------------------------------------------------

\section{Comparison To detailed numerical simulations}
\label{sec:compare 3d}

In this section we compare our model's results with a variety of
detailed numerical simulations. We show that our results are in
agreement with those of numerical simulations, with the exception of
deviations which are due to inaccuracies of the numerical calculations.

\citet{keshet2003gri} used TreeSPH simulations of LSS formation to estimate the IC
emission from electrons accelerated in LSS shocks. The
typical $>10\,\textrm{GeV}$ flux from rich clusters obtained by
\citet{keshet2003gri} is
$\textrm{few}\times10^{-7}(\eta_{e}/0.05)\,\textrm{ph}\,\textrm{cm}^{-2}\,\textrm{s}^{-1}\,\textrm{sr}^{-1}$
(see their fig.~10), consistent with the prediction of
eq.~\eqref{eq:SICshock}. \citet{keshet2003gri} also used the simulations to determine $N(>f)$, the number of sources expected with flux exceeding $f$ (at a given photon energy threshold). The model presented here for the nonthermal emission of clusters may be used to predict $N(>f)$ for a given number density of halos as function of redshift and halo mass. Such an exercise has already been carried out by \citet{waxman2000frb}, who derived analytic expressions for the $\gamma$-ray luminosity of primary electrons and used them to predict $N(>f)$ assuming a Press-Schechter \citep{presswilliam1974fga} mass function for the cluster halo density. The expressions derived here for the gamma-ray luminosity of a cluster of a given mass are identical, up to a normalization factor, to those of \citet{waxman2000frb}, implying that our prediction for $N(>f)$ would be similar to that of \citet{waxman2000frb}. Since in the normalization used by \citet{waxman2000frb} the accretion rate is $\sim3f_{\textrm{inst}}^{-1}$ larger than the one used here, and since the IC accretion luminosity is proportional to the accretion rate, the luminosity function, $N(>f)$, obtained using our normalization is related to the one derived by \citet{waxman2000frb}, $N_{\rm WL}(>f)$, by $N(>f)=N_{WL}(>3f_{\textrm{inst}}^{-1}f)$. \citet{keshet2003gri} found that their numerical source number counts, $N_{\rm num}(>f)$, fall short of the analytical prediction of \citet{waxman2000frb} by a factor of $\sim6$, in the sense that $N_{\rm num}(>f)\approx N_{WL}(>6f)$. This implies that the analysis presented here is consistent with the numerical number counts of  \citet{keshet2003gri} for $f_{\textrm{inst}}\simeq0.5$.

\begin{figure}
\epsscale{1.2} \plotone{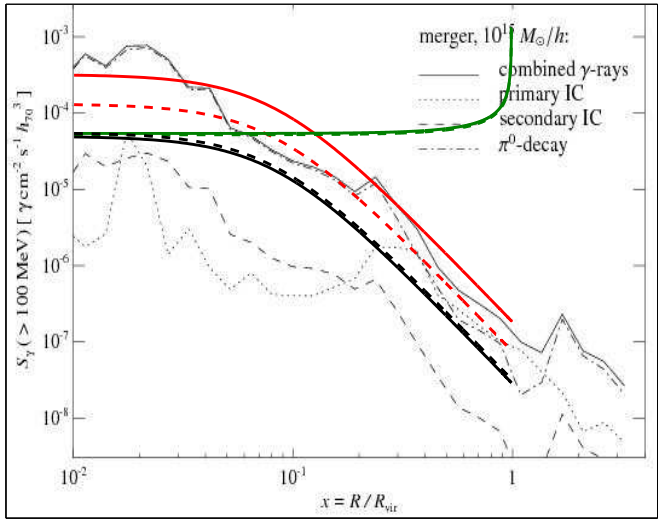} \caption{Figure $12$
of \citet{pfrommer2008scr} (curtesy of C. Pfrommer). Over plotted are the surface brightness
curves obtained from our simple analysis (eqs.~\eqref{eq:Spp},
~\eqref{eq:SICepm} and ~\eqref{eq:SICshock}), for a $T=9.4\,\textrm{keV}$,
$\beta=2/3$ cluster with $\beta_{\textrm{core}}=2\cdot10^{-3}$, $\eta_{e}=0.05$
(as chosen in the simulation), and
$f_{\textrm{inst}}=1$ ($L_{X}$ and $r_{c}$ were determined according to
eq.~\eqref{eq:power law form}).
$\beta_{\textrm{core}}=2\cdot10^{-3}$ is consistent with our model prediction,
$\beta_{\textrm{core}}\approx\eta_p/200$, for the value of $\eta_p$ chosen in
the simulation, $\eta_p=0.5$.
Line types are the same as in fig.~\ref{fig:nuLnu_total}.
Note that the rapid increase of the (primary) surface brightness near the accretion shock is likely to be suppressed by small deviations from spherical symmetry.
\label{fig:S_Pfrommer}}
\end{figure}

The results of TreeSPH simulations of LSS evolution incorporating CR generation in LSS shocks were recently reported in a series of papers \citep{pfrommer2007scr,pfrommer2008scr,pfrommer2008scrb}. In these simulations, $\eta_{p}=0.5$ and $\eta_{e}=0.05$ were adopted, and the CR evolution was followed under the assumption that the CR energy spectrum is a power-law. Figure~\ref{fig:S_Pfrommer} presents a comparison of our results with those obtained by \citet{pfrommer2008scr} for a massive ($M=10^{15}M_{\odot}$, $T=9.6\,\textrm{keV}$) merging cluster, for which detailed emission spectra are given.
The simulation's surface brightness profiles of pion decay and secondary IC emission above $100\,\textrm{MeV}$ follow the thermal surface brightness with central values of $S_{\nu}^{\textrm{pp}}(0)\simeq5\cdot10^{-4}\,\textrm{ph}\,\textrm{cm}^{-2}\,\textrm{s}^{-1}\,\textrm{sr}^{-1}$ and $S_{\nu}^{\textrm{IC},e^{\pm}}(0)\simeq5\cdot10^{-5}\,\textrm{ph}\,\textrm{cm}^{-2}\,\textrm{s}^{-1}\,\textrm{sr}^{-1}$
respectively. The profiles and the central values of the surface brightness of pion decay and secondary IC emission obtained in the simulation are consistent (to within a factor of two) with those predicted by eqs.~\eqref{eq:Spp corr} and~\eqref{eq:SICepm corr} for $\beta_{\textrm{core}}=2\cdot10^{-3}$ and $\beta=2/3$. This value of $\beta_{\textrm{core}}$ is consistent with our simple model prediction, $\beta_{\textrm{core}}\approx\eta_p/200$, for the value of $\eta_p$ chosen in the simulations, $\eta_p=0.5$. The ratio of pion decay to secondary IC luminosity obtained in the simulation is larger than predicted by our model. This may be a result of using power-law approximations for the CR emission spectra (compare the solid and dashed lines of our model near $100\,\textrm{MeV}$ in figs.~\ref{fig:nuLnu_total},~\ref{fig:S_Pfrommer}). We note that the numerical calculations, to which our model predictions are compared in fig.~\ref{fig:S_Pfrommer}, included radiative cooling of the ICM plasma, which is not included in our model. As explained in \S~\ref{sec:entropy_changes}, such radiative cooling is not expected to significantly affect the nonthermal emission of massive clusters. This conclusion is consistent with the results of the numerical calculations \citep[see, e.g., fig. 13 in][]{pfrommer2008scr}.

The value of $\beta_{\textrm{core}}$ obtained in the simulation for the cluster used for the comparison of fig.~\ref{fig:S_Pfrommer} can not be easily extracted from the results reported, since a profile of the ratio of CR to thermal gas pressure is not given. Instead, a profile of CR to thermal gas pressure ratio averaged over 9 clusters, including the one used for the comparison of fig.~\ref{fig:S_Pfrommer} and 8 smaller ($5\cdot10^{13}-5\cdot10^{14}M_\odot$) clusters, is given. This averaged pressure ratio profile is nearly constant throughout the cluster at a value of $\sim0.1$, and rises sharply within the inner $2\%$ of the virial radius to $\sim1$. A pressure ratio of $\sim0.1$ corresponds (for $\eta_p=0.5$) to $\beta_{\textrm{core}}/\eta_p\approx2/100$, larger than inferred by our analysis. However, this profile is probably biased towards larger values of $\beta_{\textrm{core}}/\eta_p$ due to the large number of small clusters for which radiative cooling is more important (radiative cooling also leads to the sharp rise at the inner $2\%$ of the virial radius). By examining the pressure ratio profiles of $M\sim10^{14}M_{\odot}$ clusters in the non-radiative simulations, which should be very similar to those of more massive clusters \citep[see also table $3$ in][]{pfrommer2007scr}, we find that indeed this ratio decreases toward the center of the cluster to a value $\beta_{\textrm{core}}/\eta_p\sim1/500$.

The primary electrons' IC surface brightness obtained in the simulation is approximately uniform across the cluster. Its value, $S_{\nu}^{\textrm{IC,shock}}\simeq5\cdot10^{-6}\,\textrm{ph}\,\textrm{cm}^{-2}\,\textrm{s}^{-1}\,\textrm{sr}^{-1}$, is lower by a factor of $\simeq10$ compared to the prediction of eq.~\eqref{eq:SICshock} (for $\beta=2/3$ and
$f_{\textrm{inst}}=1$). This discrepancy is due to the fact that the accretion shock is located at the simulation at a radius of $\sim10\,\textrm{Mpc}$, $\sim3$ times larger than $r_{200}$ (see eq.~\eqref{eq:r200,M200}), where we
assume the shock to be located. That is, the total primary IC luminosity obtained in the simulation is in agreement with our prediction, eq.~(\ref{eq:shock IC}), but the surface brightness is $\sim10$ times lower since the shock is located in the simulation at a $\sim3$ times larger radius. We believe that the shock radius is overestimated in the simulation, since our results are consistent with those of the numerical simulations of both \citet{keshet2003gri} and \citet{miniati2003nmg} (see below), which imply the shock position to be closer to $r_{200}$ than obtained by \citet{pfrommer2007scr}. Moreover, \citet{molnar2009asc} have recently shown that accretion shocks are located in SPH simulations at a radius which is $\simeq3$ larger than that obtained in   AMR simulations, probably due to numerical inaccuracies in the SPH simulations.

\begin{figure}
\epsscale{1.2} \plotone{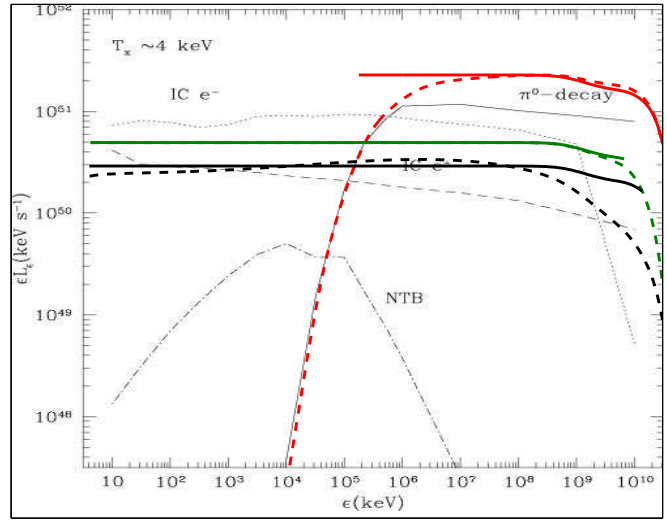} \caption{Figure $2$
from \citet{miniati2003nmg} (curtesy of F. Miniati). Over plotted are $\nu L_{\nu}$ curves
obtained from our simple analysis (eqs.~\eqref{eq:Lpp},
~\eqref{eq:sec IC} and ~\eqref{eq:shock IC}), for a $T=4\,\textrm{keV}$
and $\beta=2/3$ cluster with $\beta_{\textrm{core}}=0.05$,
$\eta_{e}=0.01$ (as chosen in the simulation), and $f_{\textrm{inst}}=1$
($L_{X}$ and $r_{c}$ were determined according to
eq.~\eqref{eq:power law form}).
The large of value of $\beta_{\textrm{core}}$ obtained in the simulations is probably
due to the low resolution of the simulation (see text). Line types
are the same as in fig.~\ref{fig:nuLnu_total}.
\label{nuLnu_miniati}}
\end{figure}

Figure~\ref{nuLnu_miniati} presents a graphical comparison of our results with those of
\citet{miniati2003nmg}, who modelled the evolution of CR protons and electrons using an Eulerian+N-body code to describe the evolution of the LSS, assuming CRs are generated at strong shocks with $\eta_{p}\sim0.6$ and $\eta_{e}=0.01$. The IC luminosity of primary electrons obtained by \citet{miniati2003nmg}, $\nu
L_{\nu}^{\textrm{IC,shock}}\simeq2\cdot10^{42}\,\textrm{erg}\,\textrm{s}^{-1}$, is
consistent with the prediction of eq.~\eqref{eq:shock IC} (for
$\beta=2/3$ and $f_{\textrm{inst}}=1$). The value of $\beta_{\textrm{core}}$ obtained in the simulation, $\beta_{\textrm{core}}\sim0.05$ for a $T=4\,\textrm{keV}$ cluster, implies a ratio $\beta_{\textrm{core}}/\eta_{p}\sim1/10$, significantly larger than the value we estimated, $\simeq1/200$, which is consistent with the results of the \citet{pfrommer2007scr} simulations. This discrepancy is most likely due to the low, $\sim100\,\textrm{kpc}$, resolution of the simulation of \citet{miniati2003nmg}, which doesn't allow one to properly resolve the adiabatic compression at the core. The pion decay luminosity and the secondaries' IC luminosity obtained by \citet{miniati2003nmg} are $\nu L_{\nu}^{\textrm{pp}}\simeq2\cdot10^{42}\,\textrm{erg}\,\textrm{s}^{-1}$
and $\nu L_{\nu}^{\textrm{IC},e^{\pm}}\simeq5\cdot10^{41}\,\textrm{erg}\,\textrm{s}^{-1}$ respectively. Both are comparable to our model predictions, eqs.~\eqref{eq:Lpp corr} and~\eqref{eq:sec IC corr}, for $T=4\,\textrm{keV}$ and $\beta_{\textrm{core}}=0.05$.

% --------------------- End of section 3  --------------------------------

% -----------------------------------------------------------------------
% --------- Sec 4: Comparison to observations - Coma cluster ------------
% -----------------------------------------------------------------------

\section{Comparison to observations - the Coma cluster}
\label{sec:compare observations}

In this section we compare the results of our analytic model to various observations of the Coma cluster. We use these observations to test our model and to constrain its parameters.
For the comparison, we use the following parameters to characterize the Coma cluster: $T_{\textrm{Coma}}=8.25\,\textrm{keV}$, $L_{\textrm{X,Coma}}=1.1\cdot10^{45}h_{70}^{-2}\,\textrm{erg}\,\textrm{s}^{-1}$, $\beta_{\textrm{Coma}}=0.654$, $r_{\textrm{c,Coma}}=246h_{70}^{-1}\,\textrm{kpc}$, $r_{200,\textrm{Coma}}\simeq2.3h_{70}^{-1}\,\textrm{Mpc}$, $M_{200,\textrm{Coma}}\simeq1.4\cdot10^{15}M_{\odot}$ and $z_{\textrm{Coma}}=0.0232$ \citep{reiprich2002mfx}. With these parameters we have $r_{\textrm{c,Coma}}/d_{\textrm{Coma}}\simeq0.14^{\circ}$ and $r_{200,\textrm{Coma}}/d_{\textrm{Coma}}\simeq1.3^{\circ}$, where $d_{\textrm{Coma}}$ is the distance to the Coma cluster. Note that the surface brightness measurements of \citet{reiprich2002mfx} reach $r_{\textrm{X,Coma}}\simeq2.9h_{70}^{-1}\,\textrm{Mpc}>r_{200,\textrm{Coma}}$, so extrapolation is not needed in order to determine $r_{200}$ and $M_{200}$ for this cluster.

In order to make the comparison with observations more straightforward, we first give below explicit expressions for the flux predicted by our model within a disk of angular radius $\theta$ (centered at the cluster's center) for a cluster at a distance $d\simeq cz/H_{0}$. We further assume that within the cluster core $B^2\gg B^2_{\textrm{CMB}}$, as inferred from radio observations \citep{kushnir2009mfc}, and normalize our results to $B_{-5}=B/10\,\mu\textrm{G}$. Using the results of \S~\ref{sec:simple}, we have
\begin{eqnarray}\label{eq:S for observations pp}
F_{\nu>\nu_{\min}}^{\textrm{pp}} (\theta) &=&
1.4\cdot10^{-11}\left(3\beta-\frac{3}{2}\right)\beta_{\textrm{core},-4}T_{1}^{1/2} \nonumber \\ &\times& \left(\frac{L_{X}}{h_{70}^{-2} 3\cdot10^{45}\,\textrm{erg}\,\textrm{s}^{-1}}\right)
 \nonumber \\
&\times& \int_{0}^{\min(\theta d/r_{c},r_{200}/r_{c})} \left(1+\bar{r}^{2}\right)^{-3\beta+1/2}\bar{r}d\bar{r}\nonumber\\
&\times&\left(\frac{\max(\varepsilon_{\nu,\min},0.1\varepsilon_{\textrm{th}})}{10\,\textrm{GeV}}\right)^{-1} \left(\frac{z}{z_{\textrm{Coma}}}\right)^{-2}\nonumber\\
 &\times&h_{70}^2 \,\textrm{ph}\,\textrm{cm}^{-2}\,\textrm{s}^{-1}
\end{eqnarray}
for $\gamma$-rays from pion decay,
\begin{eqnarray}\label{eq:S for observations secIC}
F_{\nu>\nu_{\min}}^{\textrm{IC},e^{\pm}} (\theta) &=&
3.4\cdot10^{-13}\left(3\beta-\frac{3}{2}\right)B_{-5}^{-2}\beta_{\textrm{core},-4} \nonumber \\
&\times& T_{1}^{1/2}\left(\frac{L_{X}}{h_{70}^{-2} 3\cdot10^{45}\,\textrm{erg}\,\textrm{s}^{-1}}\right) \nonumber\\
&\times& \int_{0}^{\min(\theta d/r_{c},r_{200}/r_{c})} \left(1+\bar{r}^{2}\right)^{-3\beta+1/2}\bar{r}d\bar{r} \nonumber\\
&\times& \left(\frac{\varepsilon_{\nu,\min}}{10\,\textrm{GeV}}\right)^{-1} \left(\frac{z}{z_{\textrm{Coma}}}\right)^{-2} \nonumber\\
&\times& h_{70}^2 \,\textrm{ph}\,\textrm{cm}^{-2}\,\textrm{s}^{-1}
\end{eqnarray}
for IC emission from secondaries,
\begin{eqnarray}\label{eq:S for observations secsync}
S_{\nu}^{\textrm{sync},e^{\pm}} (\theta) &=& 3.7\left(3\beta-\frac{3}{2}\right)\beta_{\textrm{core},-4}T_{1}^{1/2} \nonumber \\
&\times& \left(\frac{L_{X}}{h_{70}^{-2} 3\cdot10^{45}\,\textrm{erg}\,\textrm{s}^{-1}}\right) \nonumber\\
&\times& \int_{0}^{\min(\theta d/r_{c},r_{200}/r_{c})} \left(1+\bar{r}^{2}\right)^{-3\beta+1/2}\bar{r}d\bar{r} \nonumber\\
&\times& \left(\frac{\nu}{1.4\,\textrm{GHz}}\right)^{-1} \left(\frac{z}{z_{\textrm{Coma}}}\right)^{-2} \nonumber\\
&\times& h_{70}^2 \,\textrm{Jy}
\end{eqnarray}
for synchrotron emission from secondaries,
and
\begin{eqnarray}\label{eq:S for observations shockIC}
F_{\nu>\nu_{\min}}^{\textrm{IC,shock}}(\theta) &=&
4.7\cdot10^{-7}\left(\langle f_{\textrm{inst}}\rangle_{\theta}\eta_{e}\right)_{-2}\beta^{1/2} \nonumber \\ &\times& \left(\frac{f_{b}}{0.17}\right) T_{1}^{3/2} \left(\frac{\varepsilon_{\nu,\min}}{10\,\textrm{GeV}}\right)^{-1} \nonumber\\
&\times& g_{\rm acc.}(\theta)\bar{Z}(z) h_{70}^2(z)\,\textrm{ph}\,\textrm{cm}^{-2}\textrm{s}^{-1},
\end{eqnarray}
with
\begin{eqnarray}\label{eq:g_acc}
g_{\rm acc.}(\theta)= 2\theta_{200}^{2}\left(1-\sqrt{1-\left(\frac{\theta}{\theta_{200}}\right)^{2}}\right),
\end{eqnarray}
for IC emission of primary electrons.
Here $\theta_{200}=r_{200}/d$ and $\langle f_{\textrm{inst}}\rangle_{\theta}$ is the average value of $f_{\textrm{inst}}$ over the disk considered. Eq~(\ref{eq:g_acc}) gives an approximate description of the dependence of $F$ on $\theta$, for the case where $w\ll r_{200}$ (see eq.~\eqref{eq:xi apr}). For $w=0.1r_{200}$, eq.~(\ref{eq:g_acc}) is accurate to better than $\sim25\%$.

\subsection{HXR observations}
\label{sec:HXR}

An excess of HXR emission over the expected thermal
bremsstrahlung emission has been observed in the Coma cluster
with instruments on board three different X-ray satellites:
RXTE \citep{rephaeli1999rxr,rephaeli2002rsr}, BeppoSax \citep{fuscofemiano1999hxr,fusco2004cnh,fusco2007nhx} and INTEGRAL \citep{eckert2007swe,lutovinov2008xro}. Since the observations are in agreement with each other, we focus on the recent INTEGRAL observations. The INTEGRAL-measured flux in the $44-107\, \textrm{keV}$ band (where the thermal contribution is small) is $(1.8\pm1.1)\cdot10^{-11}\, \textrm{erg}\, \textrm{cm}^{-2}\, \textrm{s}^{-1}$ within $\theta<1^{\circ}$. To study the spatial structure of the HXR emission, images in the "soft" $17-28.5\, \textrm{keV}$ and in the "hard" $44-107\, \textrm{keV}$ INTEGRAL bands were used. In the soft band, Coma is clearly an extended source. At the hard-band, the raw image does not show any significant substructure or correlation with the cluster's thermal emission (on $1^{\circ}$ scale). The extended nature of the HXR emission implies that the radiating particles are not secondary particles produced in the interaction of cosmic-rays with the ICM, since the generation of such secondary particles should be strongly concentrated towards the cluster's center.

\begin{figure}
\epsscale{1} \plotone{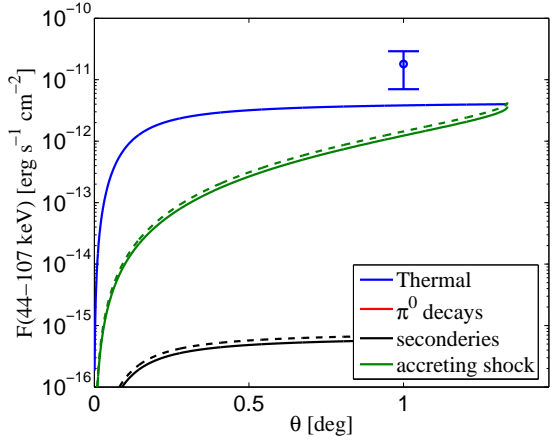} \caption{The predicted HXR flux within a disk of angular radius $\theta$ as function of $\theta$ for the Coma cluster in the $44-107\, \textrm{keV}$ band, assuming $f_{\textrm{inst}}\eta_{e}=0.01$, $\beta_{\textrm{core}}=10^{-4}$ and $B=10\,\mu\textrm{G}$. Line types are the same as in fig.~\ref{fig:nuLnu_total}. The error-bar represents INTEGRAL's measurement. According to our model, the flux is dominated by IC emission of accretion shock electrons, and INTEGRAL's measurement implies $0.06<\langle f_{\textrm{inst}}\rangle_{1^{\circ}}\eta_{e}<0.23$. The HXR excess measured by INTEGRAL can not be due to secondary emission, since $\beta_{\textrm{core}}\sim5$ would be required in order to reproduce the observed flux by secondary emission, implying that the CR energy density is much greater than the thermal energy density of the ICM.
\label{INTEGRAL_coma_total}}
\end{figure}

In fig.~\ref{INTEGRAL_coma_total} we show the estimated HXR flux within a disk of angular radius $\theta$ as function of $\theta$ for the Coma cluster in the $44-107\, \textrm{keV}$ band, assuming $f_{\textrm{inst}}\eta_{e}=0.01$, $\beta_{\textrm{core}}=10^{-4}$ and $B=10\,\mu\textrm{G}$. The INTEGRAL measurement and the thermal emission in this band are also given. Since the nonthermal flux is dominated in this band, according to our model, by IC emission of accretion shock electrons, the observed HXR flux may be used to calibrate $\eta_{e}f_{\textrm{inst}}$ ($\beta_{\textrm{core}}$ is not constrained since the contribution from secondaries is negligible). According to our model, the hard-band image should not show any significant substructure since it originates from the accretion shock, and no spatial correlation with the cluster thermal emission is expected.

Using eqs.~\eqref{eq:S for observations secIC} and \eqref{eq:S for observations shockIC}, with $\bar{Z}(0)\simeq0.96$ and
\begin{equation}\label{eq:Coma Total Integral INTEGRAL}
\int_{0}^{1^{\circ}/0.14^{\circ}}\left(1+\bar{r}^{2}\right)^{-3\beta_{\textrm{Coma}}+1/2}\bar{r}d\bar{r}\simeq0.90,
\end{equation}
we have
\begin{eqnarray}\label{eq:coma total flux INTEGERAL}
F_{\textrm{Coma,total}}(1^{\circ})&\simeq& 6.7\cdot10^{-16}\beta_{\textrm{core},-4}B_{-5}^{-2} \nonumber \\ &+& 1.4\cdot10^{-12}\left(\langle f_{\textrm{inst}}\rangle_{1^{\circ}}\eta_{e}\right)_{-2} \,\frac{\textrm{erg}}{\textrm{cm}^{2}\,\textrm{s}}
\end{eqnarray}
(where we have multiplied the number flux by $\ln(\varepsilon_{\textrm{max}}/\varepsilon_{\textrm{min}}) (1/\varepsilon_{\textrm{min}}-1/\varepsilon_{\textrm{max}})^{-1}$ with $\varepsilon_{\textrm{min}}=44\,\textrm{keV}$ and $\varepsilon_{\textrm{max}}=107\,\textrm{keV}$, in order to obtain the energy flux). The INTEGRAL measurement thus constrains $0.06<\langle f_{\textrm{inst}}\rangle_{1^{\circ}}\eta_{e}<0.23$. Note, that our analysis also rules out secondary emission as the source of the HXR excess, since in order to reproduce the observed flux by secondary emission $\beta_{\textrm{core}}\sim5$ would be required, implying that the CR energy density is much greater than the thermal energy density of the ICM.

\subsection{EGRET's $\gamma$-ray observations}
\label{sec:EGRET}

\begin{figure}
\epsscale{1} \plotone{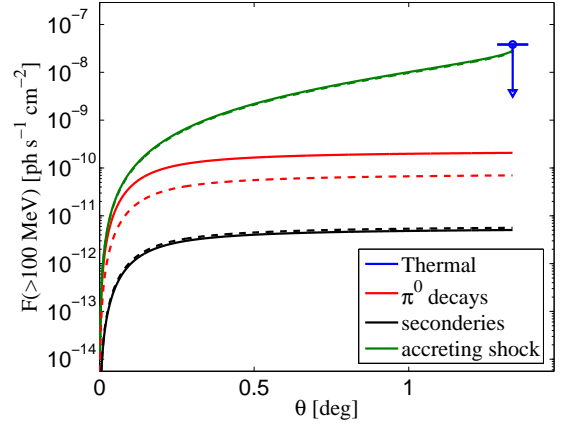} \caption{The predicted $>100\,\textrm{MeV}$ $\gamma$-ray flux within a disk of angular radius $\theta$ as function of $\theta$ for the Coma cluster, assuming $f_{\textrm{inst}}\eta_{e}=0.01$, $\beta_{\textrm{core}}=10^{-4}$ and $B=10\,\mu\textrm{G}$. Line types are the same as in fig.~\ref{fig:nuLnu_total}. The arrow represents EGRET's upper limit. The upper limit implies $\langle f_{\textrm{inst}}\rangle_{1.3^{\circ}}\eta_{e}<1.4\cdot10^{-2}$ and $\beta_{\textrm{core}}<0.05$.
\label{EGRET_coma_total}}
\end{figure}

\citet{reimer2003eul} report EGRET upper limits on high-energy $\gamma$-ray emission from the Coma cluster. EGRET's upper limit for the flux above $100\,\textrm{MeV}$ is $F_{\textrm{Coma}}<3.81\cdot10^{-8}\,\textrm{ph}\,\textrm{cm}^{-2}\,\textrm{s}^{-1}$ (with Coma assumed to be a point source; EGRET's FWHM at this energy is $5.8^{\circ}$).
In fig.~\ref{EGRET_coma_total} we compare the estimated $\gamma$-ray flux within a disk of angular radius $\theta$ as function of $\theta$ for the Coma cluster with an energy threshold of $100\,\textrm{MeV}$, assuming $f_{\textrm{inst}}\eta_{e}=0.01$, $\beta_{\textrm{core}}=10^{-4}$ and $B=10\,\mu\textrm{G}$, with EGRET's upper limit.
Since the flux is dominated in this band, according to our model, by IC emission of accretion shock electrons, EGRET's upper limit sets an upper bound for $\eta_{e}f_{\textrm{inst}}$ (and only a weak constraint on $\beta_{\textrm{core}}$; Note that the $\pi^{0}$ decays contribute significantly only above $\simeq122\,\textrm{MeV}$ due to the threshold for pion production).

Using eqs.~\eqref{eq:S for observations pp},~\eqref{eq:S for observations secIC} and~\eqref{eq:S for observations shockIC}, and
\begin{equation}\label{eq:Coma Total Integral EGRET}
\int_{0}^{r_{200,\textrm{Coma}}/r_{\textrm{c,Coma}}} \left(1+\bar{r}^{2}\right)^{-3\beta_{\textrm{Coma}}+1/2}\bar{r}d\bar{r}\simeq0.95,
\end{equation}
we have
\begin{eqnarray}\label{eq:coma total flux EGRET}
F_{\textrm{Coma,total}}&\simeq& 7.0\cdot10^{-11}\beta_{\textrm{core},-4} \nonumber \\
&+& 3.0\cdot10^{-8}\left(\langle f_{\textrm{inst}}\rangle_{1.3^{\circ}}\eta_{e}\right)_{-2}
\,\frac{\textrm{ph}}{\textrm{cm}^{2}\,\textrm{s}}
\end{eqnarray}
(since the energy band is close to the pion production threshold, we used the more detailed spectral dependence of the secondaries, as described in \S~\ref{sec:rad sum}, to obtain $F_{\textrm{Coma,pp}}\simeq7.0\cdot10^{-11}\beta_{\textrm{core},-4} \,\textrm{ph}\,\textrm{cm}^{-2}\,\textrm{s}^{-1}$ instead of $F_{\textrm{Coma,pp}}\simeq2.0\cdot10^{-10}\beta_{\textrm{core},-4} \,\textrm{ph}\,\textrm{cm}^{-2}\,\textrm{s}^{-1}$).
EGRET's upper limit implies therefore $\beta_{\textrm{core}}<0.05$ and $\langle f_{\textrm{inst}}\rangle_{1.3^{\circ}}\eta_{e}<1.4\cdot10^{-2}$. We don't regard the upper limit on $f_{\textrm{inst}}\eta_{e}$ to be in contradiction with the range calibrated in \S~\ref{sec:HXR}, since it may be explained by a spatial dependence of $f_{\textrm{inst}}$, or by a primary electron energy distribution slightly steeper than $dn/d\varepsilon\propto\varepsilon^{-2}$. Future measurements of the $\gamma$-ray emission will thus allow one to constrain the primary electron spectral index.

\subsection{VHE $\gamma$-ray observations}
\label{sec:HESS}

\begin{figure}
\epsscale{1} \plotone{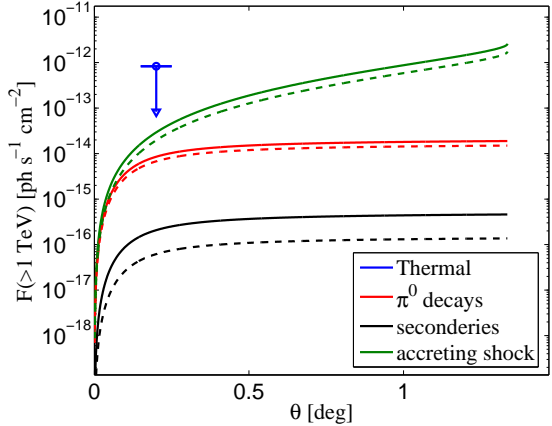} \caption{The predicted $> 1\,\textrm{TeV}$ $\gamma$-ray flux within a disk of angular radius $\theta$ as function of $\theta$ for the Coma cluster, assuming $f_{\textrm{inst}}\eta_{e}=0.01$, $\beta_{\textrm{core}}=10^{-4}$ and $B=10\,\mu\textrm{G}$. Line types are the same as in fig.~\ref{fig:nuLnu_total}.
The arrow represents  the upper limit reported by HESS \citep{domainko2007hog}, which implies $\beta_{\textrm{core}}<1.2\cdot10^{-2}$. Only weak constraints on $f_{\textrm{inst}}\eta_{e}$ may be obtained, due to the uncertainty in the maximal energy to which primary electrons may be accelerated, which is $\sim$~few~TeV.
\label{HESS_coma_total}}
\end{figure}

The Coma cluster has been observed in the VHE $\gamma$-ray band with HESS \citep{domainko2007hog} and with VERITAS \citep{perkins2008voc}. Both observation are consistent, and we focus on the HESS core observations, which provide more stringent constraints on our model. The upper limit inferred from HESS observations on the flux above $1\,\textrm{TeV}$ within $0.2^{\circ}$ is $F_{\textrm{Coma}}<8.3\cdot10^{-13}\,\textrm{ph}\,\textrm{cm}^{-2}\,\textrm{s}^{-1}$. In fig.~\ref{HESS_coma_total} we compare HESS's upper limit for the Coma cluster with the model predicted $>1\,\textrm{TeV}$ $\gamma$-ray flux (within a disk of angular radius $\theta$ as function of $\theta$), assuming $f_{\textrm{inst}}\eta_{e}=0.01$, $\beta_{\textrm{core}}=10^{-4}$ and $B=10\,\mu\textrm{G}$.
Since the energy threshold for this measurement is very high, it provides robust constraints only on $\beta_{\textrm{core}}$, through the predicted $\pi^{0}$ decay luminosity. Only weak constraints on $f_{\textrm{inst}}\eta_{e}$ may be obtained, due to the uncertainty in the maximal energy to which primary electrons may be accelerated, which is $\sim$~few~TeV.

Using eqs.~\eqref{eq:S for observations pp}, \eqref{eq:S for observations secIC} and \eqref{eq:S for observations shockIC}, and
\begin{equation}\label{eq:Coma Total Integral HESS}
\int_{0}^{0.2^{\circ}/0.14^{\circ}}\left(1+\bar{r}^{2}\right)^{ -3\beta_{\textrm{Coma}}+1/2}\bar{r}d\bar{r}\simeq0.43,
\end{equation}
we have
\begin{eqnarray}\label{eq:coma total flux HESS}
F_{\textrm{Coma,total}}&\simeq& 6.8\cdot10^{-15}\beta_{\textrm{core},-4} \nonumber \\
&+& 2.0\cdot10^{-14}\left(\langle f_{\textrm{inst}}\rangle_{0.2^{\circ}}\eta_{e}\right)_{-2}
\,\frac{\textrm{ph}}{\textrm{cm}^{2}\,\textrm{s}}
\end{eqnarray}
(due to the strong dependence of the IC flux on the maximal energy of the primary electrons, we used the exact formulae for IC emission spectra, as described in \S~\ref{sec:rad sum}).
The upper limit of HESS therefore implies $\beta_{\textrm{core}}<1.2\cdot10^{-2}$ (and $\langle f_{\textrm{inst}}\rangle_{0.2^{\circ}}\eta_{e}<0.42$, with large uncertainty).

\subsection{Radio observations}
\label{sec:radio}

\begin{figure}
\epsscale{1} \plotone{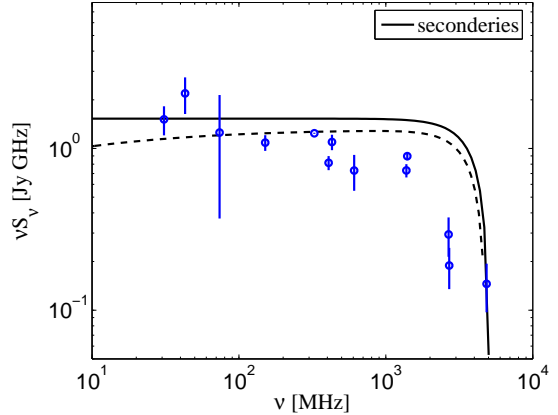} \caption{A comparison of the \citet{thierbach2003cic} compilation of flux density measurements of the Coma radio halo with our model predictions, assuming $B\gg B_{\textrm{CMB}}$, $\beta_{\textrm{core}}=2\cdot10^{-4}$, and that the radio halo flux is dominated by synchrotron emission from the secondaries (and taking into account the SZ effect). The spectral steepening observed above $\simeq2\,\textrm{GHz}$ is not robust (see text). Line types are the same as in fig.~\ref{fig:nuLnu_total}.
\label{radio_coma_total}}
\end{figure}

In fig.~\ref{radio_coma_total} we compare the \citet{thierbach2003cic} compilation of flux density measurements of the Coma radio halo with our model predictions, assuming $B\gg B_{\textrm{CMB}}$, $\beta_{\textrm{core}}=2\cdot10^{-4}$, and that the radio halo flux is dominated by synchrotron emission from the secondaries. We note that the spectral steepening observed above $\simeq2\,\textrm{GHz}$ is not robust, since it is the result of the substraction of two big numbers, the total flux and the flux of point sources, and the flux of point sources is not measured but rather extrapolated from lower frequencies assuming a constant spectral index. Since a steepening of the spectra is visible in other sources (for example, the two central galaxies of Coma: NGC4869 and NGC4874), by assuming a constant spectral index one overestimates the point source flux and underestimates the flux density of the diffuse component \citep{thierbach2003cic}. Moreover, somewhat above $\simeq3\,\textrm{GHz}$ the observed flux is suppressed also by the SZ effect \citep[see][for a detailed discussion]{ensslin2002mcr}.

Using eq.~\eqref{eq:Coma Total Integral EGRET} and eq.~\eqref{eq:S for observations secsync} we have
\begin{eqnarray}\label{eq:coma total flux radio}
&\nu S_{\nu,\textrm{Coma}}\simeq5.7\cdot10^{-1}\beta_{\textrm{core},-4}\,\textrm{Jy}\,\textrm{GHz}.
\end{eqnarray}
Comparing with observations (see fig.~\ref{radio_coma_total}) this implies $\beta_{\textrm{core}}\simeq2\cdot10^{-4}$. Note, that this constraint is consistent with the average value of $\beta_{\textrm{core}}\simeq10^{-4}$ derived by \citet{kushnir2009mfc} for a complete sample of radio emitting galaxy clusters.

% --------------------- End of section 4 --------------------------------

% -----------------------------------------------------------------------
% --------------------- Sec 5: ICM and CR evolution --------------------
% -----------------------------------------------------------------------

\section{ICM and CR evolution}
\label{sec:dynamics}

In \S~\ref{sec:simple} we derived the spectral and radial distribution of the nonthermal emission produced by ICM CRs for massive clusters, $M\gtrsim10^{14.5}M_\sun$. We assumed that the fraction $\beta_{\textrm{core}}$ of plasma energy carried by CR protons at the central regions of clusters, which dominate the emission from secondary particles, is nearly independent of cluster mass and that the scatter around its average value is small. We have argued, based on simple arguments and crude approximations, that $\beta_{\textrm{core}}\simeq\eta_{p}/100$. In what follows we present a simple model for the thermal history of the ICM, including the effects of mergers, which allows us to obtain a more accurate estimate of the value of $\beta_{\textrm{CR},p}$, its spatial dependence within clusters, and its scatter among different clusters.

We first describe in \S~\ref{sec:acc_and_merger} our model for the accretion and merger history of clusters. In \S~\ref{sec:merger} a simple model describing the effects of mergers is constructed. The results of \S~\ref{sec:acc_and_merger} and \S~\ref{sec:merger} are used in \S~\ref{sec:cr dynamics} to construct a model for the evolution of CRs in the ICM. The effects of entropy changes that are not driven by gravity (i.e. not due to accretion and merger shocks) are discussed in \S~\ref{sec:entropy_changes}. We assume throughout the analysis that the CR pressure is small compared to the thermal plasma pressure \citep[as supported by observations, see][]{kushnir2009mfc}.

\subsection{Merger and accretion history}
\label{sec:acc_and_merger}

\begin{figure}
\epsscale{1} \plotone{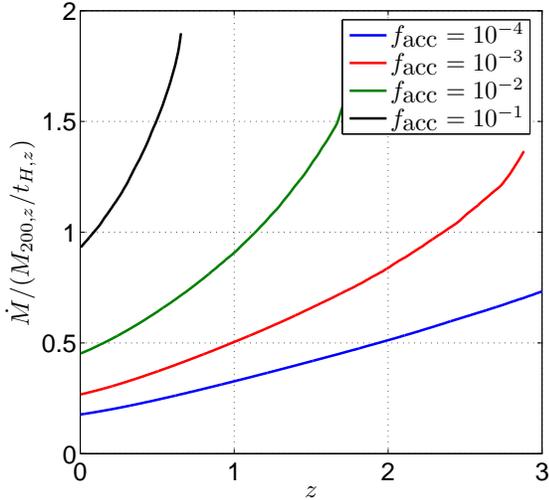} \caption{Our estimate for the accretion rate, normalized to $M_{200,z}/t_{H,z}$, for different ratios $f_{\textrm{acc}}\equiv M_{l}/M_{200,0}$. The accretion rate is averaged over $100$ realizations of a $M_{200,0}=10^{15}M_\sun$ cluster.
\label{dMdt}}
\end{figure}

We define cluster mass as the mass contained within a radius $r_{200}$, within which the mean density is $200$ times the critical density, $M_{200,z}\equiv M(r_{200})=(4/3)\pi r_{200}^{3}\times 200\rho_{\textrm{crit}}(z)$. We investigate the present, $z=0$, nonthermal emission of $M_{200,0}>10^{14.5}M_\odot$ clusters. For each value of $M_{200,0}$, chosen from a grid of masses within this range, we construct an ensemble of "merger trees" using the \citet{presswilliam1974fga} based scheme of \citet{lacey1993mrh}. In order to incorporate the accretion process into this scheme, we assume that merger events involving masses lower than a certain "resolution", $M_{l}$, are accretion events. Since the mass accretion rate, $\dot{M}$, calculated in this manner diverges within the scheme of \citet{lacey1993mrh} as the time step tends to zero, we calculate $\dot{M}$ within the frame work of Press-Schechter theory (see \S~\ref{sec:appendix B} for details). Figure~\ref{dMdt} shows $\dot{M}$ obtained in our scheme for different values of $f_{\textrm{acc}}\equiv M_{l}/M_{200,0}$, normalized to $M_{200,z}/t_{H,z}$, where $t_{H,z}$ is the Hubble time at redshift $z$. The accretion rate is averaged over $100$ realizations of a $M_{200,0}=10^{15}M_\sun$ cluster. The figure illustrates that $\dot{M}$ depends weakly on the value of $f_{\textrm{acc}}$, and is lower than $M_{200,z}/t_{H,z}$ at low redshifts by a factor of $2-5$ for mass resolutions which are adequate for a description of a merger tree (i.e. that reproduce the general results of the Press-Schechter theory).
The deviation of $\dot{M}$ from $M_{200,z}/t_{H,z}$ is expressed in terms of $f_{\textrm{inst}}\equiv \dot{M}t_{H,z}/M_{200,z}$.
As discussed in \S~\ref{sec:compare 3d}, 3D numerical simulations indicate that the average value of $f_{\textrm{inst}}$ is $\approx0.5$ at low $z$. We therefore choose  $M_{l}=2\cdot10^{-3}M_{200,0}$, which yields $f_{\textrm{inst}}\sim 0.5$ for $z<1$ and $0.5<f_{\textrm{inst}}<1$ for $1<z<2.5$. Our results depend only weakly on the value of $f_{\textrm{acc}}$, and are little modified when $f_{\textrm{acc}}=5\cdot10^{-4}$ is chosen instead of $f_{\textrm{acc}}=2\cdot10^{-3}$.

Next we describe how the gas profile is evolved with the addition of accreted mass. We assume that the halos are described, both before and after the accretion, by an NFW density profile \citep{navarro1997udp},
\begin{equation}\label{eq:NFW profile}
\rho (r)=\frac{\rho_{s}}{(r/r_{s})(1+r/r_{s})^{2}},
\end{equation}
where $\rho_{s}=M_{s}/(4\pi r_{s}^{3})$ and
\begin{equation}\label{eq:Ms def}
M_{s}=\frac{M_{200}}{ln(1+r_{200}/r_{s})-(r_{200}/r_{s})/(1+r_{200}/r_{s})}.
\end{equation}
$r_{s}$ is often expressed in terms of the "concentration parameter", $c_{200}\equiv r_{200}/r_{s}$. We adopt $c_{200}=4$ for all of our systems, a value typical of galaxy clusters simulated in a
$\Lambda$CDM concordance cosmology. Numerous studies \citep[e.g.][]{eke2001psd} found that the mass dependence of the
concentration parameter is weak, varying between different clusters by only a factor of $\sim2$.

We assume that the baryon density is proportional to the total (baryon and dark matter) density, $\rho_{\textrm{gas}}\approx f_{b}\rho$, where $f_{b}$ is the cosmic baryon fraction, in agreement with numerical simulations and observations \citep{borgani04xpg,Kravtsov05ecs,vikhlinin2006csn}. This assumption is not valid
within cluster cores, where the gas density profile is flatter than that of the dark matter. Furthermore, the NFW profile is different than the $\beta$-model we used for the gas density in \S~\ref{sec:simple}. However, the choice of an NFW or a $\beta$-model for the gas density is expected to have only a small effect on the inferred value of $\beta_{\textrm{core}}$, due to the following reason. As we show below, the production of CR protons is dominated by accretion shocks and the evolution of their energy density is affected mainly by adiabatic compression. This, combined with the fact that both the NFW and the $\beta$-model density profiles imply a ratio of $\sim100$ between the gas density in the region that dominates the secondary emission and the gas density at $r_{200}$, imply that the modification of $\beta_{\textrm{core}}$ due to the difference between the profiles of such models is small.

Another inaccuracy introduced by using an NFW gas profile is an exaggerated cooling of CR protons in the very center of clusters due to the over estimate of gas density (see eq.~\eqref{eq:p cool}). This, however, has only a minor effect on our results since the cooling affects only a small fraction of the gas mass (see \S~\ref{sec:cr dynamics}). We chose to work with an NFW density profile despite the above limitations, since it allows us to derive a very simple model for the description of cluster mergers (see \S~\ref{sec:merger}).

In order to completely specify the properties of the gas, we must choose an entropy profile. Assuming hydrostatic equilibrium,
\begin{equation}\label{eq:hydrostatic}
\frac{dp(r)}{dr}=-\frac{GM(r)}{r^{2}}\rho_{\textrm{gas}}(r),
\end{equation}
the entropy is deduced from the equation of state, $p=K\rho_{\textrm{gas}}^{5/3}$. A boundary condition must be specified for eq.~(\ref{eq:hydrostatic}). We choose to specify a value for the pressure of the gas at $r_{200}$. Although there is some freedom in the choice of the pressure at $r_{200}$, the entropy profile is rather insensitive to $p(r=r_{200})$ \citep[for physically reasonable values, see][]{mccarthy2007msh}. Since at intermediate radii the NFW profile is well approximated by an isothermal profile, i.e. $\rho\propto r^{-2}$ and $d\ln K/d\ln M_{gas}=4/3$, we choose a value of $p(r_{200})$ that establishes this near pure power-law entropy distribution all the way out to $r_{200}$. \citet{voit2002mem} have demonstrated that groups and clusters are indeed expected to have near power-law entropy distributions out to large radii.

Assuming that a hydrostatic NFW profile is preserved in the accretion process, we modify, following accretion, the density and the temperature of each mass element according to its Lagrangian location, in order to match the post-accretion hydrostatic NFW profile. We finally assume that the accreted mass was shocked at an infinite Mach number shock to the density and temperature at the outskirts of the post-accretion hydrostatic NFW profile.

\subsection{A simple model for mergers}
\label{sec:merger}

\begin{figure}
\epsscale{1} \plotone{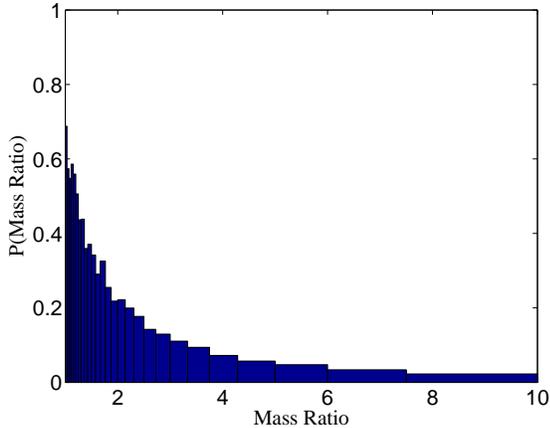} \caption{The distribution of primary to secondary halo mass ratio in mergers, obtained from $100$ realizations of the merger history for $M_{200,0}=10^{15}M_{\odot}$. The distribution is weighted by the secondary mass, since the thermodynamic variables of the secondary halo are more affected in merger events than the primary ones.
\label{MassRatio}}
\end{figure}

Let us next construct a simple model describing the evolution of ICM gas in merger events. We first note that the most frequent mergers occur between halos of similar mass. This is illustrated in fig.~\ref{MassRatio}, which shows the distribution of the ratio between the primary (heavier halo) mass, $M_p$, and secondary (lighter halo) mass, $M_s$, over $100$ runs for $M_{200,0}=10^{15}M_{\odot}$. The distribution is weighted by the secondary mass, since the thermodynamic variables of the secondary halo are more affected in merger events than the primary ones. According to fig.~\ref{MassRatio}, high mass ratio mergers, $M_p/M_s>10$, are rare. We therefore treat such mergers in a very approximated manner, as described below. Note, that although the mass accumulated in high mass ratio mergers is a small fraction of the total cluster mass, it is not necessarily small compared to the cluster core mass, which dominates the secondary emission. Thus, high mass ratio mergers could have affected the nonthermal emission, had the mass of the secondary accreted halos been accumulated in the cluster core. Numerical simulations \citep[e.g.][]{mccarthy2007msh} indicate that this is not the case: For $M_p/M_s=10$ the secondary halo penetrates through the massive one and its mass is spread, on a Hubble time scale, over the entire primary halo.

For the description of small mass ratio mergers we rely on detailed 3D numerical simulations \citep[e.g.][]{mccarthy2007msh}. We assume that the halos are described, both before and after mergers, by an NFW density profile. We show that this assumption, combined with the inference from numerical simulations that during the merger the ICM is shocked twice (with the two shocks separated by adiabatic expansion), completely determines the Mach numbers of the shocks and the magnitude of the adiabatic expansion experienced by different fluid elements.

\begin{figure}
\epsscale{1} \plotone{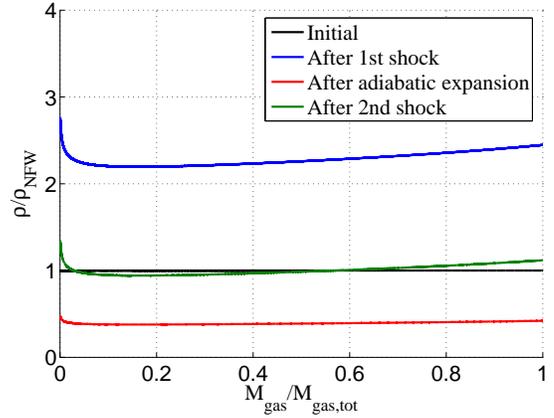} \caption{The gas density of the merged cluster as function of the accumulated gas mass for an equal mass merger, for a calibrated adiabatic expansion density decrease factor (see text) of $\simeq0.17$. The profile of the merged cluster is produced by sorting the gas elements of the initial clusters according to their density.
\label{den_profile}}
\end{figure}

\begin{figure}
\epsscale{1} \plotone{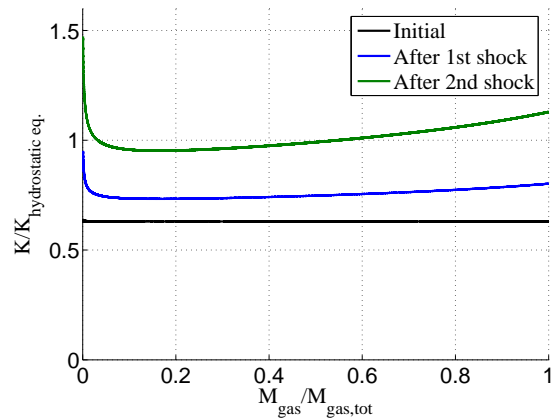} \caption{Same as fig.~\ref{den_profile} for the entropy of the merged cluster gas.
\label{ent_profile}}
\end{figure}

We assume that the available energy to be thermalized in the merger is divided equally between the two shock episodes. This determines the Mach number of each mass element in each shock episode \citep[for details, see][]{mccarthy2007msh}, such that the only two free parameters are the magnitudes of the adiabatic expansion experienced by each cluster. Assuming that the profile of the merged cluster is produced by sorting the gas elements of the initial clusters according to their density, the two adiabatic expansion factors are determined by minimizing the difference between the resulting profile and a hydrostatic NFW profile. In order to illustrate the procedure used, fig.~\ref{den_profile} and fig.~\ref{ent_profile} show the resulting density and entropy profiles for a merger with a $1:1$ mass ratio. Merging the clusters, and sorting according to density leads to an exact NFW profile, with entropy which is lower than required by hydrostatic equilibrium.
After the first shock, the entropy becomes higher but the resulting density exceeds the NFW density. Adiabatic expansion and a second shock then lead to density and entropy profiles which are very close to NFW at hydrostatic equilibrium. The expansion factor required for obtaining the required final NFW density and entropy is $\simeq6$, corresponding to a density decrease factor of $\simeq0.17$

\begin{figure}
\epsscale{1} \plotone{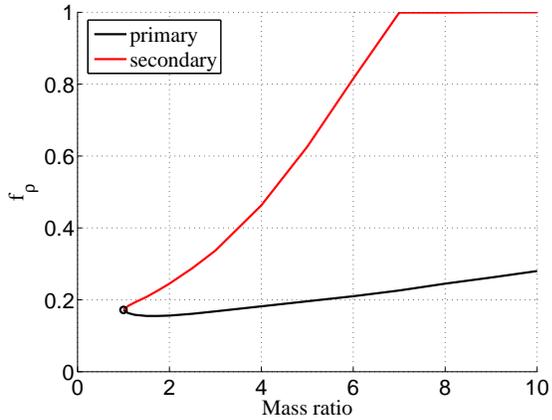} \caption{Calibrated adiabatic expansion factors as function of merging clusters mass ratio.
\label{adiabatic factors}}
\end{figure}

It is remarkable that this simple model reproduces a post-merger hydrostatic NFW halo. We note that assuming a different density profile for the halo (for example, a $\beta$-model, see \S~\ref{sec:simple}) does not allow one to preserve the density profile in the post merger cluster using such a simple model. The density decrease factors determined by this method for mergers of different mass ratios are shown in fig.~\ref{adiabatic factors}. Above a mass ratio of $10:1$ we simply sort the gas elements according to their density, and change the density and the temperature of each mass element according to its Lagrangian location, in order to reproduce the post-merger hydrostatic NFW profile. The details of large mass ratio mergers have only a small effect on   our results.

\begin{figure}
\epsscale{1} \plotone{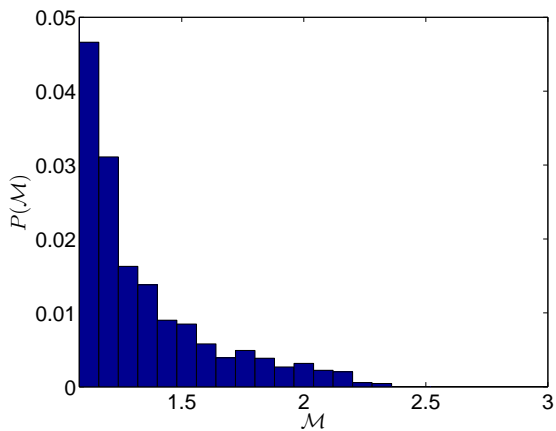} \caption{Merger shocks' Mach number distribution, weighted by the mass of the shocked gas, averaged over $100$ merger history realizations, for a cluster with $M_{200,0}=10^{15}M_{\odot}$. The first bin is not shown, since almost all of the shocks are included in it. The maximal Mach number achieved in mergers is smaller than $2.5$.
\label{Mach dist}}
\end{figure}

One of the results of our model is the merger shocks' Mach number distribution. We show this distribution, weighted by the mass of the shocked gas, in fig.~\ref{Mach dist}, based on $100$ merger history realizations for a cluster with $M_{200,0}=10^{15}M_{\odot}$ (the first bin is not shown, since almost all of the shocks are included in it). As can be seen from the figure, the maximal Mach number achieved in merger shocks is smaller than $2.5$, in agreement with the numerical simulations of \citet{skillman2008csa}. Since such weak shocks are assumed to produce very steep CR spectra (see \S~\ref{sec:Introduction}), with normalization that depends on poorly known conditions at the injection energy of the CRs, we ignore acceleration in merger shocks hereafter.

\subsection{CR evolution}
\label{sec:cr dynamics}

After determining, in \S~\ref{sec:acc_and_merger} and \S~\ref{sec:merger}, the thermodynamic history of the ICM plasma, we turn now to the evolution of the CR population. We follow here only the evolution of CR protons, since we are only interested in the value of $\beta_{\textrm{core}}$. The processes that we include in the calculations are: injection of CRs in accretion shocks (we neglect injection in weak merger shocks, see \S~\ref{sec:merger}), proton energy losses due to Coulomb and inelastic nuclear collisions, using the parametrization of \citet{kamae2006pan} for the p-p interaction, modification of the  CR energy density due to adiabatic expansion/compression (assuming the CRs to behave as a relativistic gas with an adiabatic index of $4/3$). As explained in the introduction, we assume that CR diffusion is not important, and that the CRs are coupled to the thermal plasma by magnetic fields. We also assume that the CR pressure is small compared to the thermal plasma pressure \citep[as supported by observations, see][]{kushnir2009mfc}. We calculate numerically the CR content of each mass element in the cluster, following the thermodynamic evolution of this mass element and describing the CR energy distribution using a discrete (Lagrangian, logarithmically spaced) set of energy bins.

\begin{figure}
\epsscale{1} \plotone{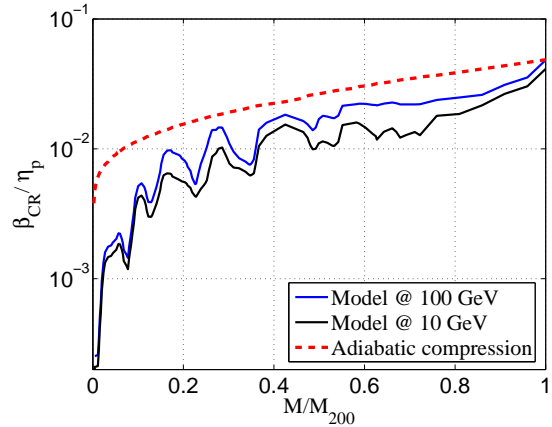} \caption{$\beta_{\textrm{CR}}/\eta_p$ at $100\,\textrm{GeV}$ (blue) and at $10\,\textrm{GeV}$ (black), as function of the accumulated mass for a typical $M_{200,0}=10^{15}M_{\odot}$ cluster. The red line shows a $\rho_{\textrm{gas}}^{-1/3}$ scaling of $\beta_{\textrm{CR}}/\eta_p$, normalized at the accretion shock.
\label{beta example}}
\end{figure}

Figure~\ref{beta example} presents typical distributions of $\beta_{\textrm{CR},p}(\varepsilon=100\,\textrm{GeV})/\eta_p$ and $\beta_{\textrm{CR},p}(\varepsilon=10\,\textrm{GeV})/\eta_p$, obtained for one realization of the evolution of a $10^{15}M_\odot$ cluster. $\beta_{\textrm{CR},p}$ depends weakly on energy, as demonstrated in the figure, from $\sim10\,\textrm{GeV}$ up to the maximal energy to which the CR protons are accelerated. The spatial distribution of $\beta_{\textrm{CR},p}$ approximately follows $\rho_{\textrm{gas}}^{-1/3}$, with normalization fixed at the accretion shock.
$\beta_{\textrm{CR},p}$ falls somewhat below this scaling at the cluster center mainly due to the effect of weak shocks (which lead to compression without significant production of high energy CRs) and due to proton energy loss at the high density cluster core. The latter effect is exaggerated in our calculations due to our assumption that the the gas density follows the dark matter density. However, cooling affects only a small fraction ($1\%$) of the gas mass. The fact that $\beta_{\textrm{CR},p}$ follows approximately $\rho_{\textrm{gas}}^{-1/3}$ implies that its evolution is determined mainly by the adiabatic compression of the ICM plasma, and that mergers do not significantly modify $\beta_{\textrm{CR},p}$.

\begin{figure}
\epsscale{1} \plotone{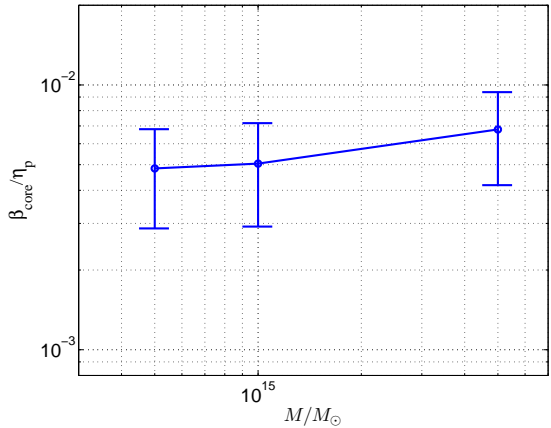} \caption{$\beta_{\textrm{core}}$ (the mass average of $\beta_{\textrm{CR},p}$ at $100\,\textrm{GeV}$ within the inner $10\%$ of the cluster mass) as function of cluster mass at $z=0$. The average value (for a given cluster mass) was obtained from $100$ realizations of merger histories, and the error bar represents the $1\sigma$ scatter of the $\beta_{\textrm{CR},p}$ distribution.
\label{beta core}}
\end{figure}

We define $\beta_{\textrm{core}}$ as the mass average of $\beta_{\textrm{CR}}$ at $100\,\textrm{GeV}$ within the inner $10\%$ of the cluster mass, which corresponds to the core region. Figure~\ref{beta core} shows the average value and the scatter of $\beta_{\textrm{core}}/\eta_p$ as function of the cluster mass at $z=0$. As can be seen from the figure, $\beta_{\textrm{core}}/\eta_p\simeq1/200$ is a good approximation for $M_{200,0}>10^{14.5}M_{\odot}$.  $\beta_{\textrm{core}}$ depends only weakly on cluster mass, changing by a factor of less than $2$ over one decade of cluster mass, with a factor $\sim2$ scatter between different clusters of given mass.

\subsection{The effects of "non-gravitational" entropy changes}
\label{sec:entropy_changes}

Our model for the ICM evolution takes into account entropy changes of the ICM driven by gravity only (accretion and merger shocks), and assumes that the galaxy clusters' density and entropy profiles are self-similar. It is well known \citep[see e.g.][]{arnaud1999tlx} that models taking into account only gravitational effects and assuming self-similarity lead to a correlation $L_{X}\propto T^{2}$ between the bolometric X-ray luminosity, $L_X$, and the temperature of the clusters, $T$, which is flatter than the observed correlation, $L_{X}\propto T^{2.5}$ \citep[see, e.g.][]{reiprich2002mfx}. This deviation from self-similarity is commonly explained as due to entropy increase of the gas at some high redshift by non-gravitational processes such as supernovae, star formation and galactic winds \citep[e.g.][]{kaiser1991ecg,evrard1991exr,navarro1995sxr,cavaliere1997tlt,balogh1999phi,ponman1999tti}. Although such entropy increase does not modify significantly the entropy profile of the massive clusters in which we are interested, one may worry about its possible effect on the CR population, since part of the gas residing in high mass clusters has, in the past, been associated with low mass clusters. The CR energy density in low mass clusters may be reduced by an early entropy increase, which reduces the Mach numbers of accretion shocks. Repeating our calculations including an early entropy increase we find, however, that this effect is rather small. Early (non-gravitational) entropy increase may reduce the Mach numbers of accretion shocks to values which significantly affect CR production only for clusters of mass $<10^{13}M_\odot$, and the suppression of CR production in such shocks reduces the the value of $\beta_{\textrm{core}}$ in massive clusters by no more than 20\%.

Another non-gravitational effect that may modify the ICM entropy is radiative cooling at the cores of massive clusters. This effect may lead to an increase of the ratio of CR to thermal ICM energy density and may therefore affect the predicted secondary emission. However, numerical simulations \citep{pfrommer2007scr} predict that the effect of cooling on the nonthermal secondary emission is not very large, as can be estimated by examining the simulated $\gamma$-ray emission from massive clusters, for radiative and non-radiative simulations. This is probably due to the fact that cooling affects mainly the inner few tens of $\textrm{kpc}$, while the secondary emission is dominated by a larger region, $\sim300\,\textrm{kpc}$. We note that it is well-known that   the effects of cooling flows are exaggerated in numerical simulations \citep[e.g.][and references therein]{balogh2001rcc,borgani04xpg,borgani2006hcb}.

% --------------------- End of section 5 --------------------------------

% -----------------------------------------------------------------------
% ---------------------------- Sec 6: Discussion ------------------------
% -----------------------------------------------------------------------

\section{Discussion}
\label{sec:discussion}

We have derived analytic expressions, eqs.~\eqref{eq:Lpp}--\eqref{eq:Ssyncshock}, which approximately describe the spectral and radial distribution of the nonthermal emission produced by (primary and secondary) CRs in massive, $M\gtrsim10^{14.5}M_\sun$, galaxy clusters. These expressions depend on the (observed) cluster thermal X-ray properties and on two model parameter, $\beta_{\textrm{core}}$ and $\eta_e$. We have shown (see \S~\ref{sec:dynamics}) that CR production is dominated by strong accretion shocks, and that the energy density of CR protons is mainly affected (after production) by the adiabatic compression of the ICM plasma. Simple arguments based on crude approximations imply that adiabatic compression should lead to $\beta_{\textrm{core}}\sim\eta_{p}/100$ \citep[\S~\ref{sec:simple}, see also][]{pfrommer2007scr,jubelgas2008crf}. A more detailed analysis of the thermodynamic history of the ICM (\S~\ref{sec:dynamics}) yields $\beta_{\textrm{core}}\simeq\eta_{p}/200$, nearly independent of cluster mass and with a scatter $\Delta\ln\beta_{\textrm{core}}\simeq1$ between clusters of given mass (see fig.~\ref{beta core}). The analysis presented in \S~\ref{sec:dynamics} includes a simple description of the thermodynamic history of the ICM, as determined by merger and accretion events, and a calculation of the energy and spatial distribution of CR protons. We have shown in \S~\ref{sec:compare observations} that our model's results agree with those of detailed numerical calculations, and that discrepancies between the results of various numerical simulations (and between such results and our model) are due to inaccuracies in the numerical calculations.

The charged secondary emission depends on the the strength of the magnetic field in the cluster core. It is given in \S~\ref{sec:sec rad} in terms of $B/B_{\textrm{CMB}}$, the ratio of the magnetic field to $B_{\textrm{CMB}}$, defined as the magnetic field for which the magnetic energy density equals the CMB energy density, $B_{\textrm{CMB}}\simeq3\,\mu \textrm{G}$. Since the secondary $e^\pm$ emission is dominated at high energy, $\gtrsim1$~eV, by secondary $\pi^0$ decay and by IC emission of primary electrons, the uncertainty in the value of $B$ affects only the predicted radio emission, which is not the main focus of the current paper. The detailed discussion of radio emission from clusters given in \citet{kushnir2009mfc} indicates that $B$ is within the range of $\sim1\mu$G to $\sim10\mu$G.

Our model predicts that the HXR and $\gamma$-ray luminosities produced by IC scattering of CMB photons by electrons accelerated in accretion shocks exceed the luminosities produced by secondary particles by factors $\simeq500(\eta_e/\eta_p)(T/10{\rm keV})^{-1/2}$ and $\simeq150(\eta_e/\eta_p)(T/10{\rm keV})^{-1/2}$ respectively, where $T$ is the cluster temperature.
Secondary particle emission may dominate at the radio and VHE ($\gtrsim1$~TeV) $\gamma$-ray bands. Our model predicts, in contrast with some earlier work that neglected the primary IC emission, that the HXR and $\gamma$-ray emission from clusters of galaxies are extended, since the emission is dominated at these energies by primary (rather than by secondary) electrons. Our model is supported by observations of the Coma cluster, where the HXR image does not show any significant substructure, the measured HXR flux corresponds to a reasonable efficiency of electron acceleration, $\eta_e\sim$~a few percent, and the radio emission corresponds to a reasonable efficiency of proton acceleration, $\beta_{\textrm{core}}\sim10^{-4}$ implying $\eta_p\sim$~a few percent   (see \S~\ref{sec:compare observations}). As explained in \S~\ref{sec:HXR} \citep[see also][]{kushnir2009mfc}, in order for secondary emission to reproduce the HXR emission of Coma, an unreasonably high CR energy density is required.

We have assumed in our analysis that the ICM plasma is isothermal and in hydrostatic equilibrium. Deviations from this simple model near the virial radius may change our estimates for $\eta_{e,p}$. Although such deviations are only weakly constrained by observations, both observational \citep[e.g.][]{vikhlinin2005ctp} and theoretical \citep[e.g.][]{roncarelli2006sxr} analyses indicated that they are not large (for example, the accretion shock temperature is lower than the virial temperature by no more than a factor $\sim2$). Modifications of the ICM properties near the virial radius may be easily incorporated into our model. Improved (observational) determination of the ICM profile near the virial radius will therefore allow one to improve the accuracy of the determination of $\eta_{e,p}$.

Our model predicts, that future HXR observations (e.g. NuStar, Simbol-X) and space-based $\gamma$-ray observations (Fermi) will lead to detection of clusters of galaxies as extended sources. Since EGRET's sensitivity is marginally too low for such detection (see \S~\ref{sec:compare observations}), we predict Fermi's $\sim50$ times better sensitivity would suffice. Single sources could be detected and resolved in both energy bands (the angular resolution of Fermi can be as low as $0.1^{\circ}$ and that of future HXR observation may reach tens of arcsec). We give in table~\ref{tbl:most IC} the predicted flux above $50\,\textrm{GeV}$ within $0.1^{\circ}$ for the $10$ most luminous clusters from the extended sample of \citet{reiprich2002mfx}, assuming $\eta_{e}f_{\rm inst}=0.01$. We also give the cluster angular radius. For cluster diameters larger than $0.2^{\circ}$ it would be possible to distinguish between extended and core emission with Fermi, thus discriminating between secondary emission from the core and primary emission from the accretion shock.

Figure~\ref{HESS_coma_total} shows that the current upper limit on the VHE $\gamma$-ray flux from the core of the Coma cluster is higher than the predicted flux by a factor of $\sim100$. This implies that a detection of the predicted Coma core flux is not possible with current generation imaging Cerenkov telescopes. Imaging telescopes may, however, allow us to detect the predicted high energy emission produced by the accretion shock. It should be pointed out that lowering the threshold energy of the imaging Cerenkov telescopes to $\sim0.1\TeV$ is important for this type of observation, since there is a large uncertainty in the predicted flux at energies $>\TeV$, due to the uncertainty in the exact value of the maximal energy to which electrons can be accelerated, which is $\sim$~few TeV. The sources in table~\ref{tbl:most IC} are also optimal for imaging telescopes.

\begin{deluxetable*}{cccrrrrrrrrrcrl}
\tablecaption{The predicted IC flux above $50\,\textrm{GeV}$ within $0.1^{\circ}$ for the $10$ most luminous clusters from the extended sample of \citet{reiprich2002mfx}, assuming $\eta_{e}f_{inst}=0.01$. \label{tbl:most IC}} \tablewidth{0pt} \tablehead{ \colhead{Cluster name} &
\colhead{Flux [$10^{-12}\,\textrm{ph}\,\textrm{cm}^{-2}\,\textrm{s}^{-1}$]} & \colhead{Angular radius [$\textrm{deg}$]}} \startdata A2163 & 0.71 & 0.21 \\
A1914 & 0.45 & 0.22 \\
Ophiuchus & 0.40 & 1.3  \\
A3888 & 0.38 & 0.25 \\
A0754 & 0.34 & 0.65 \\
A1689 & 0.33 & 0.18 \\
Triangulum & 0.31 & 0.63 \\
A2142 & 0.31 & 0.35 \\
A3266 & 0.27 & 0.56 \\
A2029 & 0.27 & 0.40 \\
\enddata

\end{deluxetable*}

In order to detect pion decay emission from cluster cores, two approaches may be adopted. One may construct high resolution ($\sim0.1^{\circ}$) HXR maps and compare them to the $\gamma$-ray maps. Since the correlation between the HXR and the $\gamma$-ray maps should be due to the primary IC emission from the accretion shock, their difference should trace the pion decay contribution. Alternatively, one may identify clusters which may be resolved and for which the surface brightness of the core is dominated by pion decays, so that a sharp decrease in the surface brightness with radius may be observed. Table~\ref{tbl:most pp} gives a list of the best candidates for such detection: The predicted flux above $50\,\textrm{GeV}$ within $0.1^{\circ}$ is given for the $10$ clusters with the highest pion to IC ratio from the extended sample of \citet{reiprich2002mfx}, assuming $\eta_{e}f_{inst}=0.01$ and $\beta_{\textrm{core}}=10^{-4}$. We considered only clusters with angular radius exceeding $0.2^{\circ}$, which may be resolved.

\begin{deluxetable*}{cccccrrrrrrrcrl}
\tablecaption{The predicted flux above $50\,\textrm{GeV}$ within $0.1^{\circ}$ for the $10$ clusters with the highest pion to IC ratio from the extended sample of \citet{reiprich2002mfx}, assuming $\eta_{e}f_{inst}=0.01$ and $\beta_{\textrm{core}}=10^{-4}$.\label{tbl:most pp}} \tablewidth{0pt} \tablehead{ \colhead{Cluster name} &
\colhead{Pion decays flux } &
\colhead{IC flux } &
\colhead{ratio} &
\colhead{Angular radius} \\
\colhead{} & \colhead{[$10^{-13}\,\textrm{ph}\,\textrm{cm}^{-2}\,\textrm{s}^{-1}$]} & \colhead{[$10^{-12}\,\textrm{ph}\,\textrm{cm}^{-2}\,\textrm{s}^{-1}$]} & \colhead{} &  \colhead{[$\textrm{deg}$]} } \startdata Perseus & 3.1 & 0.15 & 2.0 & 1.4\\
A3526 & 0.49 & 0.05 & 1.0 & 1.7\\
NGC 4636 & 0.026 & 0.003 & 0.80 & 2.2\\
2A 0335 & 0.30 & 0.038 & 0.79 & 0.50\\
NGC 5813 & 0.017 & 0.0024 & 0.73 & 1.3\\
Ophiuchus & 2.7 & 0.4 & 0.67 & 1.3\\
A0262 & 0.12 & 0.018 & 0.65 & 0.80\\
A2199 & 0.41 & 0.072 & 0.56 & 0.72 \\
A0496 & 0.33 & 0.059 & 0.56 & 0.57 \\
NGC 5044 & 0.034 & 0.0061 & 0.56 & 1.1 \\
\enddata

\end{deluxetable*}

As illustrated by the analysis of the observations of the Coma cluster, high energy and radio observations of a controlled sample of clusters will allow one to calibrate the model parameters $\eta_{e}f_{\textrm{inst}}$ and $\beta_{\rm core}$. Determination of $f_{\textrm{inst}}$ and $\beta_{\rm core}/\eta_p$ from numerical simulations would then allow one to determine $\eta_{p}$ and $\eta_{e}$. Analysis of cluster radio emission, leading to a determination of $\beta_{\rm core}$ and of the strength of magnetic fields in cluster cores, is reported by \citet{kushnir2009mfc}. Analysis of all clusters observed in HXRs, aimed at determining $\eta_{e}f_{\textrm{inst}}$, is reported by \citet{kushnir2009hxr}. Since the group of clusters observed in HXRs is not a complete controlled sample, the determination of  $\eta_{e}f_{\textrm{inst}}$ based on current observations may be biased. A controlled sample may be produced by future HXR observations (e.g. NuStar, Simbol-X). Moreover, such missions may produce high resolution HXR maps of clusters, testing the prediction that the emission is extended, i.e. originating from the accretion shock. Finally, we note that measurements of the nonthermal emission in different bands (e.g. HXR and $\gamma$-rays for electrons, radio and VHE $\gamma$-rays for protons) would allow one to constrain the energy distribution of the accelerated particles.

\acknowledgments We thank C. Pfrommer for useful discussions. This research was partially supported by ISF, Minerva and AEC grants.

% -------------------------- End of Discussion --------------------------

% -----------------------------------------------------------------------
% -------------------------------- Appendix -----------------------------
% -----------------------------------------------------------------------

\appendix

% -------------------------------- Appendix A --------------------------

\section{A. Emission mechanisms}\label{sec:appendix A}

\subsection{Thermal emission}
\label{sec:thermal}

The bolometric bremsstrahlung emissivity at some radius is given by
\begin{eqnarray}\label{eq:Lx app}
&\epsilon_{X}(r)\simeq\sqrt{\frac{8}{3\pi}}\sigma_{T}\alpha_{e}c\sqrt{m_{e}c^{2}} T^{1/2}m_{p}^{-2}\mu_{e}^{-1}
\left(\frac{1}{\mu_{H}}+\frac{4}{\mu_{\textrm{He}}}\right) \rho_{\textrm{gas}}^{2}(r),
\end{eqnarray}
(assuming the thermal Gaunt factor to be $1$ and neglecting elements heavier than Helium, which may increase the bremsstrahlung emissivity by a few tens of percent). Here $\alpha_{e}$ is the fine structure constant and we use the definitions $n_{e}=\rho_{\textrm{gas}}/(\mu_{e}m_{p})$, $n_{H}=\rho_{\textrm{gas}}/(\mu_{H}m_{p})$ and $n_{\textrm{He}}=\rho_{\textrm{gas}}/(\mu_{\textrm{He}}m_{p})$ for the electron, Hydrogen and Helium number densities, respectively (which yield $\mu_{e}\simeq1.14$, $\mu_{H}\simeq1.33$ and $\mu_{\textrm{He}}\simeq16$ for fully ionized gas with hydrogen mass fraction $\chi=0.75$).

\subsection{Neutral secondaries emission}
\label{sec:pp}

The p-p $\gamma$-ray emissivity per logarithmic photon energy bin, $\nu \epsilon_\nu^{\textrm{pp}}$, due to the decay of neutral pions produced in inelastic nuclear collisions, is given to a very good approximation by \citep{katz08}
\begin{eqnarray}\label{eq:pp app}
\nu \epsilon_{\nu}^{\textrm{pp}}(r) &\simeq&
2f_{\textrm{pp}}\cdot \varepsilon^{2}\left(\frac{dn}{d\varepsilon}\right)\cdot0.1 \frac{\rho_{\textrm{gas}}(r)}{m_{p}}\left(\frac{1}{\mu_{H}}+\frac{1}{\mu_{\textrm{He}}}\right) c\sigma_{\textrm{pp}}^{\textrm{inel}} \nonumber\\
&\simeq&0.2f_{\textrm{pp}}\frac{3}{2} c\beta_{\textrm{CR}}m_{p}^{-2} \mu^{-1}\left(\frac{1}{\mu_{H}}+\frac{1}{\mu_{\textrm{He}}}\right) T\rho_{\textrm{gas}}^{2}(r)\sigma_{\textrm{pp}}^{\textrm{inel}},
\end{eqnarray}
where $\sigma_{\textrm{pp}}^{\textrm{inel}}\simeq40\,\textrm{mb}$ and $f_{\textrm{pp}}\simeq0.75$. Since the CR protons have flat spectra, $\nu \epsilon_\nu^{\textrm{pp}}$ is not energy dependent at the relevant photon energies, from $\sim0.1\,\textrm{GeV}$ ($\varepsilon_{\textrm{th}}\simeq1.22\,\textrm{GeV}$ is the threshold energy for pion production and the photon energy is $\sim0.1$ of the proton energy) up to the cutoff energy due to pair production in interaction with the IR background, which should be around $\sim10\,\textrm{TeV}$ for nearby clusters \citep{franceschini2008eoi}. Using eq.~\eqref{eq:Lx app} and eq.~\eqref{eq:pp app}, the total p-p $\gamma$-ray luminosity, $\nu L_{\nu}^{\textrm{pp}}$, is related to the bolometric bremsstrahlung luminosity, $L_{X}$, by
\begin{eqnarray}\label{eq:Lpp app}
\nu L_{\nu}^{\textrm{pp}}&\simeq&\frac{\mu_{e}}{\mu} \frac{\frac{1}{\mu_{H}}+\frac{1}{\mu_{\textrm{He}}}}{\frac{1}{\mu_{H}}+\frac{4}{\mu_{\textrm{He}}}}
\frac{\frac{3}{2}\cdot0.2f_{\textrm{pp}} \sigma_{\textrm{pp}}^{\textrm{inel}}\beta_{\textrm{CR}}}{\sqrt{\frac{8}{3\pi}}\sigma_{T}\alpha_{e}\sqrt{m_{e}c^{2}}} T^{1/2}L_{X} \nonumber \\
&\simeq& 1.3\cdot10^{41}\beta_{\textrm{core},-4}T_{1}^{1/2}\left(\frac{L_{X}}{h_{70}^{-2} 3\cdot10^{45}\,\textrm{erg}\,\textrm{s}^{-1}}\right)\,\textrm{erg}\,\textrm{s}^{-1}.
\end{eqnarray}
Using the correlations eq.~\eqref{eq:power law form}, we have
\begin{eqnarray}\label{eq:Lpp corr app}
\nu L_{\nu}^{\textrm{pp}}\simeq 1.2\cdot10^{41}\beta_{\textrm{core},-4}T_{1}^{3.06}\,\textrm{erg}\,\textrm{s}^{-1}.
\end{eqnarray}

The p-p $\gamma$-ray surface brightness above some energy $\varepsilon_{\nu,\textrm{min}}$ at some distance $\bar{r}\equiv r/r_{c}$ from the cluster center is given by \citep[for $\beta>0.5$, see][]{Sarazin1977xle}
\begin{eqnarray}\label{eq:Spp app}
S_{\nu>\nu_{\textrm{min}}}^{\textrm{pp}} (\bar{r})&\simeq&  \frac{1}{2\pi^{5/2}}0.886\nu L_{\nu}^{\textrm{pp}} r_{c}^{-2}\left(3\beta-\frac{3}{2}\right) \left(1+\bar{r}^{2}\right)^{-3\beta+1/2} \left(\max(\varepsilon_{\nu,\textrm{min}},0.1\varepsilon_{\textrm{th}})\right)^{-1} \nonumber\\
&\simeq& 5.5\cdot10^{-7}\left(3\beta-\frac{3}{2}\right)\beta_{\textrm{core},-4}T_{1}^{1/2} \left(\frac{L_{X}}{h_{70}^{-2} 3\cdot10^{45}\,\textrm{erg}\,\textrm{s}^{-1}}\right) \left(\frac{r_{c}}{h_{70}^{-1}\,200\,\textrm{kpc}}\right)^{-2}
 \nonumber\\ &\times&
\left(\frac{\max(\varepsilon_{\nu,\textrm{min}},0.1\varepsilon_{\textrm{th}})}{10\,\textrm{GeV}}\right)^{-1}
\left(1+\bar{r}^{2}\right)^{-3\beta+1/2}
\,\textrm{ph}\,\textrm{cm}^{-2}\,\textrm{s}^{-1}\textrm{sr}^{-1}
\end{eqnarray}
(neglecting the $\gamma\gamma\rightarrow e^{+}e^{-}$ cutoff). Using the correlations eq.~\eqref{eq:power law form} we have
\begin{eqnarray}\label{eq:Spp corr app}
&S_{\nu>\nu_{\textrm{min}}}^{\textrm{pp}} (\bar{r})\simeq 4.1\cdot10^{-7}\left(3\beta-\frac{3}{2}\right)\beta_{\textrm{core},-4}T_{1}^{0.42} \left(\frac{\max(\varepsilon_{\nu,\textrm{min}},0.1\varepsilon_{\textrm{th}})}{10\,\textrm{GeV}}\right)^{-1}
\left(1+\bar{r}^{2}\right)^{-3\beta+1/2} \,\textrm{ph}\,\textrm{cm}^{-2}\,\textrm{s}^{-1}\,\textrm{sr}^{-1}.
\end{eqnarray}

\subsection{IC emission from charged secondaries}
\label{sec:sec IC}

p-p collisions also produce secondary electrons and positrons, which cool by emitting synchrotron radiation and by IC scattering of CMB photons. We assume that the distribution of the secondaries is in a steady state, where in the relevant energy bands the secondaries that are generated lose all their energy to radiation. Since the CR protons have flat spectra, the secondary production ($\varepsilon L_{\varepsilon}^{e^{\pm}}$) in the range $0.1\varepsilon_{\textrm{th}}<\varepsilon_{e^{\pm}}<0.1\varepsilon_{\textrm{max}}$ is not energy dependent. To a very good approximation, the production in this energy range satisfies
\begin{equation}\label{eq:sec production app}
\varepsilon L_{\varepsilon}^{e^{\pm}}\simeq\frac{f_{e^{-}}+f_{e^{+}}}{4}\nu L_\nu^{\textrm{pp}},
\end{equation}
where $f_{e^{-}}\simeq0.8$ and $f_{e^{+}}\simeq1.2$.
As explained in \S~\ref{sec:simple}, we consider only secondaries with cooling time shorter than $t_{\rm dyn}$. For values $1-10\,\mu \textrm{G}$ of the magnetic field, the Lorentz factor of secondaries with cooling time which equals the dynamical time is $200\lesssim\gamma_{\textrm{cool}}\lesssim2000$.

To a very good approximation, in the range of photon energies $\gamma_{\textrm{cool}}^{2}3T_{\textrm{CMB}}(1+z)^4<\varepsilon_{\textrm{ph}} <\gamma_{\textrm{max}}^{2}3T_{\textrm{CMB}}(1+z)^4$, where $\gamma_{\textrm{max}}\simeq0.1\varepsilon_{\textrm{max}}/m_{e}c^{2}$ ($\varepsilon_{\textrm{max}}$ is the maximal energy of the CR protons and the secondaries energy is $\sim0.1$ of the proton energy), the luminosity due to IC scattering from the secondaries is given by
\begin{eqnarray}\label{eq:sec IC app}
\nu L_{\nu}^{\textrm{IC},e^{\pm}}&\simeq&
\frac{1}{2}\varepsilon L_{\varepsilon}^{e^{\pm}}\frac{B_{\textrm{CMB}}^{2}}{B_{\textrm{CMB}}^{2}+B^{2}}\simeq
\frac{f_{e^{+}}+f_{e^{-}}}{8}\nu L_{\nu}^{\textrm{pp}}\frac{B_{\textrm{CMB}}^{2}}{B_{\textrm{CMB}}^{2}+B^{2}} \nonumber\\
&\simeq& 3.3\cdot10^{40}\frac{B_{\textrm{CMB}}^{2}}{B_{\textrm{CMB}}^{2}+B^{2}} \beta_{\textrm{core},-4}T_{1}^{1/2}\left(\frac{L_{X}}{h_{70}^{-2} 3\cdot10^{45}\,\textrm{erg}\,\textrm{s}^{-1}}\right)\,\textrm{erg}\,\textrm{s}^{-1},
\end{eqnarray}
and the surface brightness is given by
\begin{eqnarray}\label{eq:SICepm app}
S_{\nu>\nu_{\textrm{min}}}^{\textrm{IC},e^{\pm}} (\bar{r}) &\simeq& \frac{1}{2\pi^{5/2}}0.886\nu L_{\nu}^{\textrm{IC},e^{\pm}} r_{c}^{-2}\left(3\beta-\frac{3}{2}\right) \left(1+\bar{r}^{2}\right)^{-3\beta+1/2} \left(\varepsilon_{\nu,\textrm{min}}\right)^{-1}\nonumber\\
&\simeq&1.4\cdot10^{-7}\left(3\beta-\frac{3}{2}\right)\frac{B_{\textrm{CMB}}^{2}}{B_{\textrm{CMB}}^{2}+B^{2}} \beta_{\textrm{core},-4} T_{1}^{1/2} \left(\frac{L_{X}}{h_{70}^{-2} 3\cdot10^{45}\,\textrm{erg}\,\textrm{s}^{-1}}\right) \left(\frac{r_{c}}{h_{70}^{-1}\,200\,\textrm{kpc}}\right)^{-2}
 \nonumber\\ &\times&
\left(\frac{\varepsilon_{\nu,\textrm{min}}}{10\,\textrm{GeV}}\right)^{-1} \left(1+\bar{r}^{2}\right)^{-3\beta+1/2} \,\textrm{ph}\,\textrm{cm}^{-2}\, \textrm{s}^{-1}\, \textrm{sr}^{-1}.
\end{eqnarray}

Using the correlations eq.~\eqref{eq:power law form}, we have
\begin{eqnarray}\label{eq:sec IC corr app}
&\nu L_{\nu}^{\textrm{IC},e^{\pm}}\simeq 3.0\cdot10^{40}\frac{B_{\textrm{CMB}}^{2}}{B_{\textrm{CMB}}^{2}+B^{2}} \beta_{\textrm{core},-4}T_{1}^{3.06}\,\textrm{erg}\,\textrm{s}^{-1},
\end{eqnarray}
and
\begin{eqnarray}\label{eq:SICepm corr app}
&S_{\nu>\nu_{\textrm{min}}}^{\textrm{IC},e^{\pm}} (\bar{r})\simeq 1.0\cdot10^{-7}\left(3\beta-\frac{3}{2}\right)\frac{B_{\textrm{CMB}}^{2}}{B_{\textrm{CMB}}^{2}+B^{2}} \beta_{\textrm{core},-4} T_{1}^{0.42}
\left(\frac{\varepsilon_{\nu,\textrm{min}}}{10\,\textrm{GeV}}\right)^{-1}
\left(1+\bar{r}^{2}\right)^{-3\beta+1/2} \,\textrm{ph}\,\textrm{cm}^{-2}\, \textrm{s}^{-1}\, \textrm{sr}^{-1}.
\end{eqnarray}

\subsection{Synchrotron emission from charged secondaries}
\label{sec:sec sync}

To a very good approximation, in the range of photon energies $\gamma_{\textrm{cool}}^{2}\varepsilon_{0}<\varepsilon_{\textrm{ph}}<\gamma_{\textrm{max}}^{2}\varepsilon_{0}$, where $\varepsilon_{0}=3heB/4\pi m_{e}c$, the synchrotron luminosity from the secondaries is given by
\begin{eqnarray}\label{eq:sec sync app}
\nu L_{\nu}^{\textrm{sync},e^{\pm}}&\simeq& \nu L_{\nu}^{\textrm{IC},e^{\pm}}\frac{B^{2}}{B_{\textrm{CMB}}^{2}} \nonumber\\
&\simeq& 3.3\cdot10^{40}\frac{B^{2}}{B_{\textrm{CMB}}^{2}+B^{2}} \beta_{\textrm{core},-4} T_{1}^{1/2} \left(\frac{L_{X}}{h_{70}^{-2} 3\cdot10^{45}\,\textrm{erg}\,\textrm{s}^{-1}}\right) (1+z)^{-4}\, \,\textrm{erg}\,\textrm{s}^{-1}.
\end{eqnarray}
and the surface brightness is given by:
\begin{eqnarray}\label{eq:Ssyncepm app}
S_{\nu}^{\textrm{sync},e^{\pm}} (\bar{r})&\simeq& \frac{1}{2\pi^{5/2}}0.886\nu L_{\nu}^{\textrm{sync},e^{\pm}} r_{c}^{-2}\left(3\beta-\frac{3}{2}\right) \left(1+\bar{r}^{2}\right)^{-3\beta+1/2} \nu^{-1}\nonumber\\
&\simeq& 13\left(3\beta-\frac{3}{2}\right)\frac{B^{2}}{B_{\textrm{CMB}}^{2}+B^{2}} \beta_{\textrm{core},-4}T_{1}^{1/2} \left(\frac{L_{X}}{h_{70}^{-2} 3\cdot10^{45}\,\textrm{erg}\,\textrm{s}^{-1}}\right) \left(\frac{r_{c}}{h_{70}^{-1}\,200\,\textrm{kpc}}\right)^{-2}
 \nonumber\\ &\times& \left(\frac{\nu}{1.4\,\textrm{GHz}}\right)^{-1} \left(1+\bar{r}^{2}\right)^{-3\beta+1/2}
 (1+z)^{-4}\,\textrm{mJy}\,\textrm{arcmin}^{-2}.
\end{eqnarray}

Using the correlations eq.~\eqref{eq:power law form}, we have
\begin{eqnarray}\label{eq:sec sync corr app}
&\nu L_{\nu}^{\textrm{sync},e^{\pm}}\simeq 3.0\cdot10^{40}\frac{B^{2}}{B_{\textrm{CMB}}^{2}+B^{2}} \beta_{\textrm{core},-4} T_{1}^{3.06}
(1+z)^{-4}\,\textrm{erg}\,\textrm{s}^{-1}
\end{eqnarray}
and
\begin{eqnarray}\label{eq:Ssyncepm corr app}
S_{\nu}^{\textrm{sync},e^{\pm}} (\bar{r})&\simeq&
9.8\left(3\beta-\frac{3}{2}\right)\frac{B^{2}}{B_{\textrm{CMB}}^{2}+B^{2}} \beta_{\textrm{core},-4}T_{1}^{0.42}
 \nonumber \\
&\times& \left(\frac{\nu}{1.4\,\textrm{GHz}}\right)^{-1} \left(1+\bar{r}^{2}\right)^{-3\beta+1/2} (1+z)^{-4}\,\textrm{mJy}\,\textrm{arcmin}^{-2}.
\end{eqnarray}

\subsection{Primary electrons IC emission}
\label{sec:e IC rad}

As in the case of the charged secondaries, we assume that the distribution of the primaries is in a steady state, since electrons (at the relevant energies) lose all their energy to radiation on a time scale short compared to the cluster dynamical time, $t_{\rm dyn}\sim1\,\textrm{Gyr}$. Unlike the secondary electrons and positrons, which lose energy through both IC and synchrotron emission, primary electrons lose their energy mainly by IC scattering of CMB photons, since the magnetic field at the accretion shock is expected to be weak, $\sim0.1\,\mu {\rm G}\ll B_{\rm CMB}$ \citep{waxman2000frb}. Thus the steady state assumption holds for primaries with Lorenz factors $\gamma>\gamma_{\textrm{cool}}\sim2000$.

To a very good approximation, in the range of photon energies $\gamma_{\textrm{cool}}^{2}3T_{\textrm{CMB}}(1+z)^4<\varepsilon_{\textrm{ph}} <\gamma_{\textrm{max}}^{2}3T_{\textrm{CMB}}(1+z)^4$, the luminosity of IC emission by primary, shock accelerated, electrons is given by
\begin{eqnarray}\label{eq:shock IC app}
\nu L_{\nu}^{\textrm{IC,shock}} &\simeq&
\frac{1}{2}\frac{3}{2}\frac{\eta_{e}f_{b}(\mu m_{p})^{-1}T}{\ln(p_{\textrm{max}})-\ln(m_{e}c)}f_{\textrm{inst}}\frac{M_{200}}{t_H} \nonumber\\
&\simeq& 1.8\cdot10^{43}\left(f_{\textrm{inst}}\eta_{e}\right)_{-2}\beta^{3/2}\left(\frac{f_{b}}{0.17}\right) T_{1}^{5/2}\bar{Z}(z) \,\textrm{erg}\,\textrm{s}^{-1},
\end{eqnarray}
where $\left(f_{\textrm{inst}}\eta_{e}\right)_{-2}\simeq f_{\textrm{inst}}\eta_{e}/10^{-2}$, $f_{b}=\Omega_{b}/\Omega_{m}$ and $\bar{Z}(z)\equiv(t_{H}H(z))^{-1}$. Assuming that the primary electron emission originates from a thin layer of thickness $w$ behind the accretion shock, the IC surface brightness is given by
\begin{eqnarray}\label{eq:SICshock app}
S_{\nu>\nu_{\textrm{min}}}^{\textrm{IC,shock}}(r)&\simeq& \frac{1}{8\pi^{2}r_{200}^{2}} \xi\left(r/r_{200},w/r_{200}\right) \nu L_{\nu}^{\textrm{IC,shock}} \left(\varepsilon_{\nu,\textrm{min}}\right)^{-1} \nonumber\\
&\simeq&1.5\cdot10^{-7}\left(f_{\textrm{inst}}\eta_{e}\right)_{-2}\beta^{1/2}\left(\frac{f_{b}}{0.17}\right) T_{1}^{3/2} \nonumber\\
&\times& \left(\frac{\varepsilon_{\nu,\textrm{min}}}{10\,\textrm{GeV}}\right)^{-1} \xi\left(r/r_{200},w/r_{200}\right) \nonumber\\
&\times& \bar{Z}(z)  h_{70}^2(z) \,\textrm{ph}\,\textrm{cm}^{-2}\, \textrm{s}^{-1}\, \textrm{sr}^{-1},
\end{eqnarray}
where
\begin{eqnarray}\label{eq:xi def app}
\xi(x,y)=\left\{%
\begin{array}{ll}
    \frac{3\left(\sqrt{1-x^{2}}-\sqrt{(1-y)^{2}-x^{2}}\right)}{1-\left(1-y\right)^{3}}, & \hbox{$x\leq1-y$} \\
    \frac{3\sqrt{1-x^{2}}}{1-\left(1-y\right)^{3}}, & \hbox{$1-y<x<1$} \\
\end{array}%
\right. .
\end{eqnarray}
For $w\ll r_{200}$, $\xi(x,y)$ can be approximated in the regime $x<1-y$ as
\begin{eqnarray}\label{eq:xi apr app}
\xi(x,y)\simeq\frac{1}{\sqrt{1-x^{2}}}.
\end{eqnarray}
The thickness of the emitting region is approximately given by the product of the cooling time of the emitting electrons and the velocity of the downstream fluid relative to the shock velocity. As explained in \S~\ref{sec:simple}, the approximation $w\ll r_{200}$ holds for $\gamma>\gamma_{\textrm{cool}}\sim2000$.

\subsection{Primary electrons synchrotron emission}
\label{sec:e sync rad}

The synchrotron luminosity from these electrons in the range of photon energies $\gamma_{\textrm{cool}}^{2}\varepsilon_{0}<\varepsilon_{\textrm{ph}}<\gamma_{\textrm{max}}^{2}\varepsilon_{0}$, is given by
\begin{eqnarray}\label{eq:shock sync app}
\nu L_{\nu}^{\textrm{sync,shock}} &\simeq& \nu L_{\nu}^{\textrm{IC,shock}}\frac{B^2}{B_{\rm CMB}^{2}} \nonumber\\
&\simeq&1.7\cdot10^{40}\left(f_{\textrm{inst}}\eta_{e}\right)_{-2}\beta^{3/2}\left(\frac{f_{b}}{0.17}\right) T_{1}^{5/2}B_{-7}^{2}
\bar{Z}(z) (1+z)^{-4} \,\textrm{erg}\,\textrm{s}^{-1},
\end{eqnarray}
where $B_{-7}=B/0.1\,\mu\textrm{G}$.
The synchrotron surface brightness from the shock is
\begin{eqnarray}\label{eq:Ssyncshock app}
S_{\nu}^{\textrm{sync,shock}}(r)&\simeq& \frac{1}{8\pi^{2}r_{200}^{2}} \xi\left(r/r_{200},w/r_{200}\right) \nu L_{\nu}^{\textrm{sync,shock}} \left(\nu\right)^{-1} \nonumber \\
&\simeq&1.4\cdot10^{-2}\left(f_{\textrm{inst}}\eta_{e}\right)_{-2}\beta^{1/2}\left(\frac{f_{b}}{0.17}\right) T_{1}^{3/2}B_{-7}^{2}
\left(\frac{\nu}{1.4\,\textrm{GHz}}\right)^{-1} \nonumber\\
&\times&\xi\left(r/r_{200},w/r_{200}\right)\bar{Z}(z)  h_{70}^2(z) (1+z)^{-4}\,\textrm{mJy}\,\textrm{arcmin}^{-2}.
\end{eqnarray}

% -------------------------------- End of appendix A -------------------

% -------------------------------- Appendix B --------------------------

\section{B. Press-Schechter accretion rate}\label{sec:appendix B}

In the Press-Schechter formalism, it is convenient to use
$\omega\equiv\delta_{c}(z)=\delta_{c,0}/D(z)$ as the time variable
(where $\delta_{c,0}$ is the density contrast estimated by linear
theory for the virialization of a spherical halo, and $D(z)$ is the
linear growth factor), and $S(M)\equiv\sigma^{2}(M)$, the mass
variance, as the mass variable. The probability for a step
$\Delta S$ in a time-step $\Delta\omega$ is
\begin{equation}\label{eq:numerical mass function}
P(\Delta S,\Delta\omega)d\Delta
S=\frac{1}{\sqrt{2\pi}}\frac{\Delta\omega}{(\Delta
S)^{3/2}}\exp\left[-\frac{(\Delta\omega)^{2}}{2\Delta
S}\right]d\Delta S.
\end{equation}
Changing variables,
$x\equiv\Delta\omega/(2\sqrt{\Delta S})$, this becomes a Gaussian
distribution in $x$ with zero mean and unit variance. Since $\Delta S\propto \Delta\omega^{2}$, the accretion rate, $\Delta S/\Delta\omega$, is linear in $\Delta\omega$ and thus tends to zero as $\Delta\omega\rightarrow 0$.

In order to obtain a finite mass accretion rate in our scheme, we introduce a slight correction to the treatment of masses below the mass resolution, $M_{l}$: Each time a mass step $\Delta M$ is drawn, which is below the mass resolution $M_{l}$, we assume that mass was accreted during $\Delta\omega$ at a constant rate,  $\left(dM/d\omega\right)_{\textrm{M}_{l},\textrm{acc}}$, which is the average accretion rate below $M_{l}$. Let us next calculate $\left(dM/d\omega\right)_{\textrm{M}_{l},\textrm{acc}}$. The average mass of the progenitors of some mass $M_{0}$ that contributed to $M_{0}$ through accretion over a time step $\Delta\omega$ is
\begin{equation}\label{eq:average progenitor}
\bar{M}_{\textrm{acc},\Delta\omega}=\int^{M_{0}}_{M_{0}-M_{l}}P(M|M_{0},\Delta\omega)MdM=
M_{0}\left[1-\textrm{erf}\left(\frac{\Delta\omega}{\sqrt{2S(M_{l})-2S(M_{0})}}\right)\right].
\end{equation}
Since
\begin{equation}\label{eq:aaccretion rate1}
\frac{\bar{M}_{\textrm{acc},\Delta\omega}-M_{0}}{\Delta\omega}\xra[\Delta\omega\rightarrow0]{}
-\sqrt{\frac{2}{\pi}}\frac{M_{0}}{\sqrt{S(M_{l})-S(M_{0})}},
\end{equation}
we have
\begin{equation}\label{eq:aaccretion rate2}
\left(\frac{dM}{d\omega}\right)_{\textrm{M}_{l},\textrm{acc}}=
-\sqrt{\frac{2}{\pi}}\frac{M_{0}}{\sqrt{S(M_{l})-S(M_{0})}}.
\end{equation}

% -------------------------------- End of appendix B -------------------

%-----------------------------------------------------------------------------
% --------------------------      BIBLIOGRAPGHY ---------------------------
%-----------------------------------------------------------------------------

%\clearpage

\bibliography{ms}
\bibliographystyle{hapj}

% ------------------------------ End of bibliography --------------------

\end{document}